\documentclass[aps,pra,twocolumn,superscriptaddress,groupedaddress]{revtex4}
\usepackage{epsfig}
\usepackage{hyperref}
\usepackage{braket}
\usepackage{bm,color}
\usepackage{amssymb}
\usepackage{enumitem}
\usepackage{float,CJK}

\makeatletter
\@addtoreset{equation}{section}
\@addtoreset{figure}{section}
\@addtoreset{table}{section}
\makeatother

\def\be{\begin{equation}}
\def\ee{\end{equation}}
\def\ba{\begin{eqnarray}}
\def\ea{\end{eqnarray}}
\def\sdg{Schr\"odinger}
\def\mh{\mathcal H}
\def\E{Everett~}
\begin{document}
\title{Everett's Theory of the Universal Wave Function}
\begin{CJK}{UTF8}{gbsn}
\author{Biao Wu(吴飙)}
\affiliation{International Center for Quantum Materials, School of Physics, 
Peking University, 100871, Beijing, China}
\affiliation{Wilczek Quantum Center, School of Physics and Astronomy, 
Shanghai Jiao Tong University, Shanghai 200240, China}
\affiliation{Collaborative Innovation Center of Quantum Matter, Beijing 100871,  China}

\date{\today}


\begin{abstract}
This is a tutorial for the many-worlds theory by Everett, which includes some of my personal views. 
It has two main parts.The first main part shows the emergence of many worlds in a universe consisting of only a Mach-Zehnder interferometer. 
The second main part is an abridgment of Everett's long thesis, where his 
theory was originally elaborated in detail with clarity and rigor.  Some minor comments
are added in the abridgment in light of  recent developments. Even if you do not agree to Everett's view, you will still learn a great deal 
from his generalization of the uncertainty relation, his unique way of defining entanglement (or canonical correlation), 
his formulation of quantum measurement using Hamiltonian, and his relative state. 
\end{abstract}
\maketitle
\end{CJK}




\part{Prologue}
Although Everett's many-worlds theory is now well known, it is still a minority 
view among physicists. There are many reasons, which I have no intension to discuss extensively here. 
One of the reasons may be that many physicists have not read his work seriously.  
Everett's theory was presented in his PhD thesis, which has two versions. The long version has
over 130 pages and was finished 
in 1956. It was published only 17 years later for the first time with the title {\it The Theory of 
The Universal Wave Function} in the book edited by DeWitt and Graham~\cite{DeWitt} and 
was re-published with commentary in 2012~\cite{Byrne2}.
Due to Bohr's objection, \E had to shorten it. The short version became 
his official PhD thesis at Princeton University~\cite{Byrne} and 
was published with the title {\it ``Relative State" Formulation of Quantum Mechanics} in 
Review of Modern Physics~\cite{Everett} accompanied by an article by his advisor Wheeler~\cite{Wheeler}.\\

On the one hand, Everett's short thesis lacks many important results in his long thesis, e.g., entanglement (or canonical correlation) 
and formulation of quantum measurement. On the other hand, the long version 
may be too long for many people's patience, which is further exasperated by Everett's mathematical notations 
that are not familiar to modern readers. It is my hope that this abridgment 
makes a good compromise between the long and short thesis. 
In this abridged version, I will  keep its structure and stick to Everett's original statements as much as possible at key points 
while omitting detailed discussion and derivations. 
Entanglement is all over the long thesis. However, \E never used the word entanglement; instead, he called it
canonical correlation or simply correlation. I will use entanglement in this abridgment. 
In addition, I'll use Dirac brackets wherever possible.  \\

Before the abridgment, I use the Mach-Zehnder interferometer (MZI) to illustrate  the many-worlds theory. 
It appears to me that the MZI is a simple example to illustrate all the essential points  in Everett's long thesis. 
In particular,   the MZI is  ideal to demonstrate interference between different worlds and  the essence of approximate measurement, which was 
discussed in detail by \E in his long thesis and has not been discussed much since.  Near the end of this part, 
I also discuss the issue of preferred basis and I think that it is related to the perceptive abilities of observers. 
Hopefully, this  example of MZI will aid your reading of the abridgment. \\

\begin{figure}[h]
 \includegraphics[width=7.5cm]{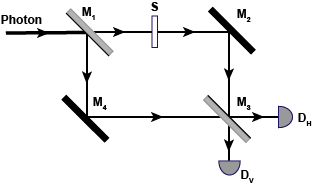}
        \caption{Mach-Zehnder interferometer. $M_1$ and $M_3$ are half-silvered mirrors; $M_2$ and $M_4$ are reflective mirrors. 
        $S$ is some sample material that causes a phase shift to the photon's wave function. $D_H$ and $D_V$ are two photon detectors. 
        Note that $M_1$ and $M_3$ are not symmetric and their two sides are made of materials of different refractive indices.}
	\label{mzi}
\end{figure}

The two words, universe and world, are often used differently by different people when they discuss 
Everett's theory. It was first dubbed ``many-worlds" theory by DeWitt~\cite{DeWitt}. In this way, 
we say that there is one universe that consists of many different worlds. 
However,  Everett's theory has recently often been called  the theory of multiverse.  
In this way, we say that there is one world that consists of many different universes~\cite{Deutsch}.  It rubs salt to the wound 
that multiverse has different meanings for different people in literature~\cite{Kragh,Tegmark}. So, to avoid the confusion,
we stick to DeWitt's term and say that there is one universe that consists of many different worlds. \\

It is  my sincere hope that you will eventually find time to read Everett's long thesis in its entirety, which is richer in content than
the short version and juicier than this abridgment. Finally, even if you do not agree to his view, you will certainly get entertained and inspired 
by how \E generalized uncertainty relation, 
defined entanglement (or canonical correlation),  formulated quantum measurement, and introduced relative state. 

\part{The universe of MZI}
The Mach-Zehnder interferometer (MZI) was proposed by Zehnder in 1891~\cite{Zehnder} and was refined by Mach in 1892~\cite{Mach}. 
Due to its simplicity and flexibility,  the MZI has not only enjoyed wide applications~\cite{Hariharan} but also 
often been used for illustration of  fundamental subtleties in quantum mechanics~\cite{Griffiths,Deutsch}. We follow the crowd 
and use it to illustrate Everett's many-worlds theory. We first briefly review the Mach-Zehnder interferometer (MZI) 
in a conventional way.  

As shown in Fig.\ref{mzi}, 
MZI consists of four mirrors. The mirrors $M_1$ and $M_3$ are half-silvered and serve as beam-splitter. 
The mirrors $M_2$ and $M_4$ are reflective. Initially the photon's wave function  
has only the horizontal component, i.e., $\ket{\psi_0}=\ket{\phi_H}$.  After the photon encounters  $M_1$, 
its wave function splits and has two components
\be
\label{m01}
\ket{\psi_1}=\frac{1}{\sqrt{2}}(\ket{\phi_H}-\ket{\phi_V})\,.
\ee
Note that there is an ambiguity for the phase difference between the transmitted component $\ket{\phi_H}$ and the reflection component $\ket{\phi_V}$.
In general, the phase difference depends on the incident side of the beam splitter. If $\delta_1$ is the phase difference when the photon is incident on
the left side of the beam splitter and $\delta_2$ is the phase difference when the photon is incident on
the right side, then $\delta_1+\delta_2=\pi$~\cite{Zeilinger}.  The exact values of $\delta_1$ and $\delta_2$ depend on how the beam splitter 
is manufactured. In the above  we have  used $\delta_1=\pi$ and $\delta_2=0$. This occurs when the two sides of the beam splitter are made of  
materials of different refractive indices~\cite{Zetie}. With this choice of phase differences, the beam splitter functions exactly as a Hadamard gate. \\

The sample $S$ causes a phase shift $\theta$ to the wave function that passes through; the reflections 
by the mirrors $M_2$ and $M_4$ interchange $\ket{\phi_H}$ and $\ket{\phi_V}$ and 
cause a $\pi$ phase shift. As a result, right before encountering the mirror $M_3$, the photon's wave function becomes
\be
\ket{\psi_1^\prime}=\frac{1}{\sqrt{2}}(e^{i\theta}\ket{\phi_H}-\ket{\phi_V})\,.
\ee
At the mirror $M_3$, $\ket{\phi_H}$ further splits into two components becomes $(\ket{\phi_H}+\ket{\phi_V})/\sqrt{2}$ and 
$\ket{\phi_V}$  splits into $(-\ket{\phi_H}+\ket{\phi_V})/\sqrt{2}$. Consequently, we  have
\ba
\ket{\psi_2}&=&\frac{1}{2}\big[e^{i\theta}(\ket{\phi_H}+\ket{\phi_V})+\ket{\phi_H}-\ket{\phi_V}\big]\nonumber\\
&=&e^{i\theta/2}\big(\cos\frac{\theta}{2}\ket{\phi_H}+i\sin\frac{\theta}{2}\ket{\phi_V}\big)\,.
\label{emzi}
\ea
This means that the probability that the photon be detected by the detector $D_H$ is $\cos^2(\theta/2)$ and 
the probability detected by the detector $D_V$ is $\sin^2(\theta/2)$. \\

The mysterious part of the MZI is the following. On the one hand,  the photon wave function has 
two parts, $\ket{\phi_H}$ and $\ket{\phi_V}$,  right before the detection. On the other hand, in a single run of the experiment, 
there is only one detection either at $D_H$ or $D_V$. Suppose that  $D_H$ detects a  photon in one experiment; 
this detection is clearly triggered by  the $\ket{\phi_H}$  term in Eq.(\ref{emzi}). 
So, why does not the other term $\ket{\phi_V}$  trigger a detection at $D_V$? What has happened to $\ket{\phi_V}$?
According to the conventional view ,  both $\ket{\phi_H}$ and $\ket{\phi_V}$ can trigger detection, 
but it is purely random which one triggers. Furthermore, when one of them triggers detection, the other part magically 
disappears.  This is called the collapse of wave function.
\E showed in details the  collapse of wave function would lead to two difficulties  in his long thesis~\cite{DeWitt}.  
The first difficulty is that it would lead to logical inconsistency when there are two or more observers; the second difficulty
is that it is inadequate to deal with approximate measurement. \\

In both his long thesis and short thesis~\cite{Everett,DeWitt}, \E had used branches or just elements of superposition instead of worlds referring to the different superposition components in a wave function. Here in this work, we will often use  worlds 
as his theory is now widely known as the many-worlds theory~\cite{DeWitt}. \\

We now analyze MZI with the many-worlds theory. We assume that the universe consists only of MZI and nothing else. 
There is no gravity.  The two detectors can absorb the photon with 100\% efficiency. 
The mirrors $M_1$ and $M_3$ are at rest initially and arranged as in Fig.\ref{mzi} with no support or 
attached wires while  both the mirrors $M_2$ and $M_4$ are fixed in space.  With mirrors $M_1$ and $M_3$ movable, 
we can discuss the conditions for interference to occur. 
If one consider a more complicated situation where neither of the mirrors $M_2$ and $M_4$ are fixed,  the analysis would
become much more complicated without gaining essentially new physics. \\

There are two different kinds of interactions in this universe of MZI: photon with half-silvered mirror and photon with the reflective mirror. 
We use $U_e$ denote the former and $U_0$  the latter.  The interaction at the mirror $M_1$ can be mathematically
expressed as 
\ba
&&U_e\ket{\phi_H}\otimes \ket{\psi^{M_1}_0}\nonumber\\
&=&\frac{1}{\sqrt{2}}\left(\ket{\phi_H}\otimes \ket{\psi^{M_1}_0}-\ket{\phi_V}\otimes \ket{\psi^{M_1}_{p_1}}\right)\,,
\label{m1}
\ea
where $\ket{\psi^{M_1}_0}$ and $\ket{\psi^{M_1}_{p_1}}$ are the states of the mirror before 
and after the interaction, respectively. After the interaction, if the photon continues
to move horizontally, nothing changes; if the photon moves vertically, the mirror acquires a momentum $\vec{p}_1$ and 
its state becomes $\ket{\psi^{M_1}_{p_1}}$. Overall,  it results an entangled state between the photon and the mirror. 
Similarly, at the mirror $M_3$, we have 
\ba
&&U_e\ket{\phi_H}\otimes \ket{\psi^{M_3}_0}\nonumber\\
&=&\frac{1}{\sqrt{2}}\left(\ket{\phi_H}\otimes \ket{\psi^{M_3}_0}+\ket{\phi_V}\otimes \ket{\psi^{M_3}_{p_1}}\right)\,,
\label{m31}
\ea
and 
\ba
&&U_e\ket{\phi_V}\otimes \ket{\psi^{M_3}_0}\nonumber\\
&=&\frac{1}{\sqrt{2}}\left(\ket{\phi_V}\otimes \ket{\psi^{M_3}_0}-\ket{\phi_H}\otimes \ket{\psi^{M_3}_{p_3}}\right)\,.
\label{m33}
\ea
Note that $\vec{p}_1=-\vec{p}_3$. The reflective interaction at the mirror $M_2$ has the following mathematical form
\be
U_0\ket{\phi_H}\otimes \ket{\psi^{M_2}_0}=-\ket{\phi_V}\otimes \ket{\psi^{M_2}_{0}}\,.
\ee 
And similarly at the mirror $M_4$, we have
\be
U_0\ket{\phi_V}\otimes \ket{\psi^{M_4}_0}=-\ket{\phi_H}\otimes \ket{\psi^{M_4}_{0}}\,.
\ee 
No entanglement is generated in this interaction and the mirrors do not gain momentum as they are fixed in space. \\

Initially the universe of MZI is described by the following wave function
\ba
&&\ket{\Psi_0}=\ket{\phi_H}\otimes\ket{\psi^{M_1}_0}\otimes\ket{\psi^{M_2}_0}\nonumber\\
&&\otimes\ket{\psi^{M_3}_0}\otimes\ket{\psi^{M_4}_0}\otimes\ket{\psi^{D_H}_0}\otimes\ket{\psi^{D_V}_0}\,.
\ea
Whenever there is no confusion arising, we omit $\otimes$ and simplify the above the expression as
\be
\ket{\Psi_0}=\ket{\phi_H,\psi^{M_1}_0,\psi^{M_2}_0,\psi^{M_3}_0,\psi^{M_4}_0,\psi^{D_H}_0,\psi^{D_V}_0}\,.
\ee
After the photon interacts with the mirror $M_1$, we have 
\ba
\ket{\Psi_1}&=&U_e\ket{\Psi_0}
=\frac{1}{\sqrt{2}}\left(\ket{\phi_H,\psi^{M_1}_0}-\ket{\phi_V,\psi^{M_1}_{p_1}}\right)\nonumber\\
&&\otimes\ket{\psi^{M_2}_0,\psi^{M_3}_0,\psi^{M_4}_0,\psi^{D_H}_0,\psi^{D_V}_0}\,.
\ea
According to the many-worlds theory, the two components in $\ket{\Psi_1}$, which are orthogonal to each other,  
represent two different worlds: in one world the photon travels horizontally and in the other world 
the photon travels vertically. After the sample $S$, we still have two worlds but one world has acquired a phase shift
\ba
\ket{\Psi_1^\prime}
&=&\frac{1}{\sqrt{2}}\left(e^{i\theta}\ket{\phi_H,\psi^{M_1}_0}-\ket{\phi_V,\psi^{M_1}_{p_1}}\right)\nonumber\\
&&\otimes\ket{\psi^{M_2}_0,\psi^{M_3}_0,\psi^{M_4}_0,\psi^{D_H}_0,\psi^{D_V}_0}\,.
\ea
The photon is then reflected by the two mirrors $M_2$ and $M_4$ and the state of the universe becomes
\ba
\ket{\Psi_2}&=&U_0\ket{\Psi_1^\prime}
=\frac{1}{\sqrt{2}}\left(\ket{\phi_H,\psi^{M_1}_{p_1}}-e^{i\theta}\ket{\phi_V,\psi^{M_1}_0}\right)\nonumber\\
&&\otimes\ket{\psi^{M_2}_{0},\psi^{M_3}_0,\psi^{M_4}_{0},\psi^{D_H}_0,\psi^{D_V}_0}\,.
\ea
The universe still has only two worlds. Now the photon interact with the mirror $M_3$, resulting  the following state of the universe
\ba
\ket{\Psi_3}&=&U_e\ket{\Psi_2}\nonumber\\
&=&\frac{1}{2}\Big[\ket{\phi_H,\psi^{M_1}_{p_1},\psi^{M_3}_0}+\ket{\phi_V,\psi^{M_1}_{p_1},\psi^{M_3}_{p_1}}\nonumber\\
&&-e^{i\theta}(\ket{\phi_V,\psi^{M_1}_0,\psi^{M_3}_0}-\ket{\phi_H,\psi^{M_1}_0,\psi^{M_3}_{p_3}})\Big]\nonumber\\
&&\otimes\ket{\psi^{M_2}_{0},\psi^{M_4}_{0},\psi^{D_H}_0,\psi^{D_V}_0}\,.
\ea
It appears that there are now four worlds in the universe. But in general the four terms above are not orthogonal to each other.
There are in fact seven worlds. To see it,  let us expand  $\ket{\psi^{M_1}_{p_1}}$ as 
\be
\label{exp1}
\ket{\psi^{M_1}_{p_1}}=\alpha_1\ket{\psi^{M_1}_0}+\beta_1\ket{\psi^{M_1}_\perp}\,,
\ee  
where $\braket{\psi^{M_1}_\perp|\psi^{M_1}_0}=0$.  Similarly, we have 
\be
\label{exp2}
\ket{\psi^{M_3}_{p_1}}=\alpha_3\ket{\psi^{M_3}_0}+\beta_3\ket{\psi^{M_3}_\perp}\,,
\ee  
and 
\be
\label{exp3}
\ket{\psi^{M_3}_{p_3}}=\alpha_3^*\ket{\psi^{M_3}_0}+\beta_3^*\ket{\psi^{M_3*}_\perp}\,,
\ee  
where $\braket{\psi^{M_3}_\perp|\psi^{M_3}_0}=\braket{\psi^{M_3*}_\perp|\psi^{M_3}_0}=0$. In the above
we have used that $\vec{p}_1=-\vec{p}_3$ implies $\ket{\psi^{M_3}_{p_1}}=\ket{\psi^{M_3*}_{p_3}}$.
We will discuss these coefficients $\alpha_{1,3}$ and $\beta_{1,3}$ later.
With these expansions, we have 
\ba
\ket{\Psi_3}&=&\frac{1}{2}\Big[(\alpha_1+e^{i\theta}\alpha_3^*)\ket{\phi_H,\psi^{M_1}_0,\psi^{M_3}_0}\nonumber\\
&&+(\alpha_1\alpha_3-e^{i\theta})\ket{\phi_V,\psi^{M_1}_0,\psi^{M_3}_0}\nonumber\\
&&+\beta_1\ket{\phi_H,\psi^{M_1}_\perp,\psi^{M_3}_0}+\beta_3^*e^{i\theta}\ket{\phi_H,\psi^{M_1}_0,\psi^{M_3*}_\perp}\nonumber\\
&&+\alpha_3\beta_1\ket{\phi_V,\psi^{M_1}_\perp,\psi^{M_3}_0}+\alpha_1\beta_3\ket{\phi_V,\psi^{M_1}_0,\psi^{M_3}_\perp}\nonumber\\
&&+\beta_1\beta_3\ket{\phi_V,\psi^{M_1}_\perp,\psi^{M_3}_\perp}\Big]\nonumber\\
&&\otimes\ket{\psi^{M_2}_{0},\psi^{M_4}_{0},\psi^{D_H}_0,\psi^{D_V}_0}\,.
\ea
These seven terms are orthogonal to each other. In the end,  the photon is  detected by the detectors and 
we have a universe that consists of  seven different worlds that exists simultaneously. And the wave function of the universe is
\ba
\ket{\Psi_f}&=&\frac{1}{2}\Big[(\alpha_1+e^{i\theta}\alpha_3^*)\ket{\phi_H,\psi^{M_1}_0,\psi^{M_3}_0,\psi^{D_H}_1,\psi^{D_V}_0}\nonumber\\
&&+(\alpha_1\alpha_3-e^{i\theta})\ket{\phi_V,\psi^{M_1}_0,\psi^{M_3}_0,\psi^{D_H}_0,\psi^{D_V}_1}\nonumber\\
&&+\beta_1\ket{\phi_H,\psi^{M_1}_\perp,\psi^{M_3}_0,\psi^{D_H}_1,\psi^{D_V}_0}\nonumber\\
&&+\beta_3^*e^{i\theta}\ket{\phi_H,\psi^{M_1}_0,\psi^{M_3*}_\perp,\psi^{D_H}_1,\psi^{D_V}_0}\nonumber\\
&&+\alpha_3\beta_1\ket{\phi_V,\psi^{M_1}_\perp,\psi^{M_3}_0,\psi^{D_H}_0,\psi^{D_V}_1}\nonumber\\
&&+\alpha_1\beta_3\ket{\phi_V,\psi^{M_1}_0,\psi^{M_3}_\perp,\psi^{D_H}_0,\psi^{D_V}_1}\nonumber\\
&&+\beta_1\beta_3\ket{\phi_V,\psi^{M_1}_\perp,\psi^{M_3}_\perp,\psi^{D_H}_0,\psi^{D_V}_1}\Big]\nonumber\\
&&\otimes\ket{\psi^{M_2}_{0},\psi^{M_4}_{0}}\,.
\ea
In this final state, everything in the universe except the mirrors $M_2$ and $M_4$ are entangled together. 
The photon detection should be presented as 
\be
\ket{\phi_H}\otimes \ket{\psi^{D_H}_0}\rightarrow \ket{\psi^{D_H}_1}
\ee
to reflect the fact that the photon is absorbed by the detector. However, in the above, to explicitly represent the photon state
before the detection, we have kept $\phi_H$ and $\phi_V$. This should not cause confusion. \\

\begin{figure}[t]
 \includegraphics[width=7.5cm]{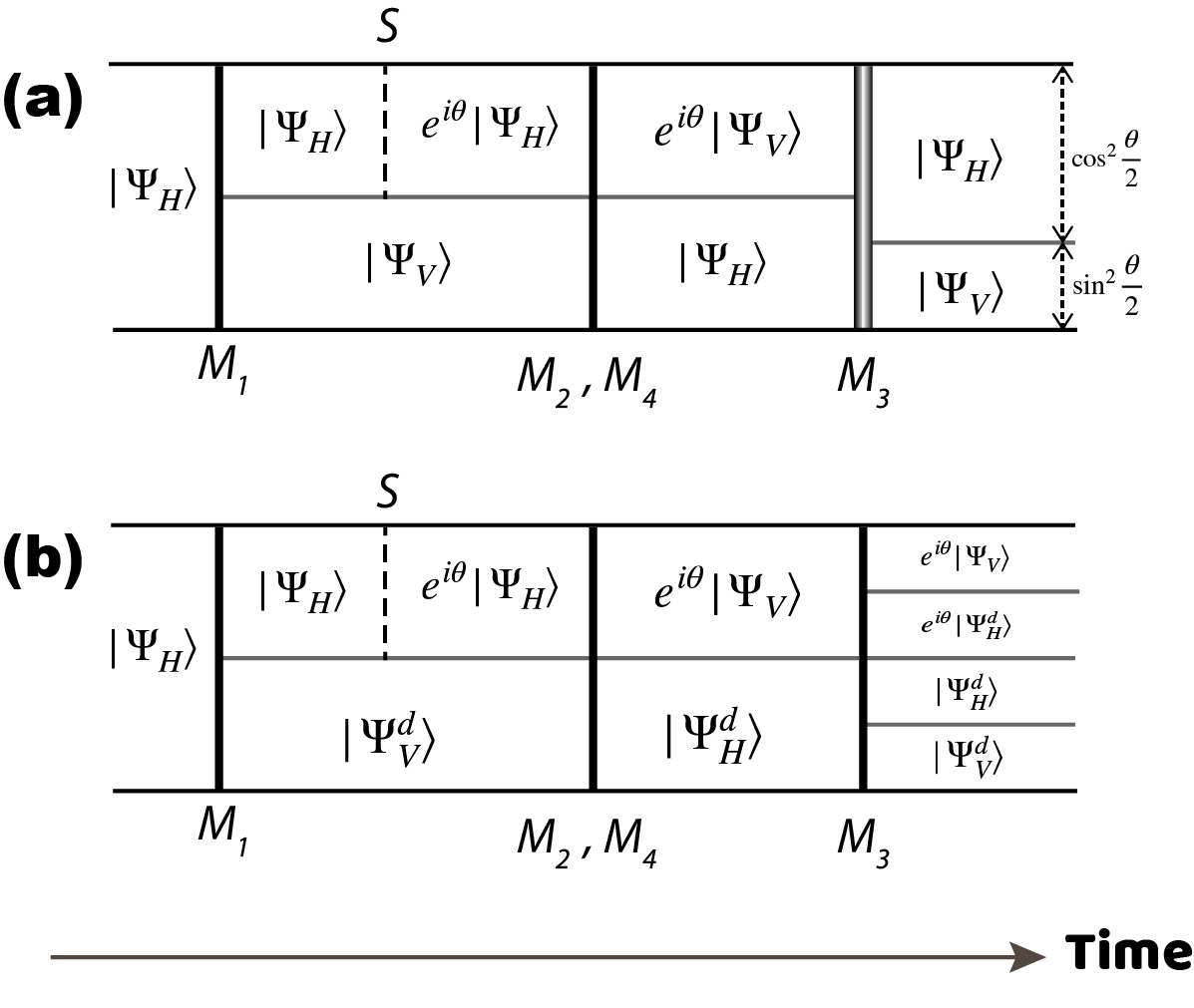}
        \caption{Worlds in the universe of Mach-Zehnder interferometer in two special cases. (a) The pure interference (PI) case; (b) the 
        pure split (PS) case. $\ket{\Psi_H}$ and $\ket{\Psi_V}$ represent worlds where the photon moves horizontally and vertically, respectively.
         The superscript $d$ in $\ket{\Psi_H^d}$ and $\ket{\Psi_V^d}$ represents the states of mirrors
         has changed relative to the initial state of the MZI universe. }
	\label{mw}
\end{figure}

We consider two special cases. In the first case,  which we call pure interference (PI) case, 
$\alpha_1=\alpha_3= 1$ and $\beta_1=\beta_3= 0$.  In the PI case, we have 
 \ba
\ket{\Psi_{f1}}&=&\frac{1}{2}\Big[(1+e^{i\theta})\ket{\phi_H,\psi^{M_1}_0,\psi^{M_3}_0,\psi^{D_H}_1,\psi^{D_V}_0}\nonumber\\
&&+(1-e^{i\theta})\ket{\phi_V,\psi^{M_1}_0,\psi^{M_3}_0,\psi^{D_H}_0,\psi^{D_V}_1}\Big]\nonumber\\
&&\otimes\ket{\psi^{M_2}_p,\psi^{M_4}_p}\nonumber\\
&=&e^{i\theta/2}\Big[\cos\frac{\theta}{2}\ket{\phi_H,\psi^{D_H}_1,\psi^{D_V}_0}\nonumber\\
&&+\sin\frac{\theta}{2}\ket{\phi_V,\psi^{D_H}_0,\psi^{D_V}_1}\Big]\nonumber\\
&&\otimes\ket{\psi^{M_1}_0,\psi^{M_3}_0,\psi^{M_2}_0,\psi^{M_4}_0}\,.
\ea
This is exactly  the state in Eq.(\ref{emzi}). The only difference is that the states of mirrors and detectors are not expressed explicitly 
in Eq.(\ref{emzi}).  This case happens when the two mirrors $M_1$ and $M_3$ are very massive or mounted in space and unmovable. 
Fig.\ref{mw}(a) shows how the worlds split and evolve in this case. Initially, there is only one world
and it splits into two worlds with equal weight at the mirror $M_1$. These two worlds evolve in parallel without changing their weights 
before interfering at the mirror $M_3$. As a result of the interference,  we still have two worlds but with different weights. 
In the special case $\theta=0$,  there is only one world after the interference. \\

Consider the second special case, $\alpha_1=\alpha_3= 0$ and $\beta_1=\beta_3= 1$.  We call it pure split (PS) case. In the PS case, we have
\ba
\ket{\Psi_{f2}}&=&\frac{1}{2}\Big[\ket{\phi_H,\psi^{M_1}_{p_1},\psi^{M_3}_0,\psi^{D_H}_1,\psi^{D_V}_0}\nonumber\\
&&+e^{i\theta}\ket{\phi_H,\psi^{M_1}_0,\psi^{M_3}_{p_3},\psi^{D_H}_1,\psi^{D_V}_0}\nonumber\\
&&-e^{i\theta}\ket{\phi_V,\psi^{M_1}_0,\psi^{M_3}_0,\psi^{D_H}_0,\psi^{D_V}_1}\nonumber\\
&&+\ket{\phi_V,\psi^{M_1}_{p_1},\psi^{M_3}_{p_1},\psi^{D_H}_0,\psi^{D_V}_1}\Big]\nonumber\\
&&\otimes\ket{\psi^{M_2}_0,\psi^{M_4}_0}\,.
\ea
The evolution of the worlds in this case is illustrated in Fig.\ref{mw}(b). The evolution is similar to the first case before 
the mirror $M_3$. The crucial difference is that there is no interference at $M_3$ in this case. Consequently, the worlds keep splitting 
and we obtain four different worlds with equal weights. And the phase shift $\theta$ has no effect 
on the weights of the different worlds. \\

We now examine the expansions in Eqs.(\ref{exp1},\ref{exp2},\ref{exp3}) in detail. The mirrors, made of atoms, have enormous amount 
of degrees of freedom. However, in this MZI universe,  only their centers of mass are relevant. Moreover, their centers of mass 
move only along $\vec{p}_1=-\vec{p}_3$. With these considerations, we are allowed to describe the states of the mirror $M_1$ 
before and after the interaction as the following Gaussian wave packets
\be
\braket{x|\psi^{M_1}_{0}}=\left(\frac{1}{\pi a^2}\right)^{1/4}\exp\Big(-\frac{x^2}{2a^2}\Big)\,,
\ee
and 
\be
\braket{x|\psi^{M_1}_{p_1}}=\left(\frac{1}{\pi a^2}\right)^{1/4}\exp\Big(-\frac{x^2}{2a^2}+ikx\Big)\,,
\ee
where $a$ is the width of the wave packet and $\hbar k=|\vec{p}_1|$.  We obtain
\be
\alpha_1=\braket{\psi^{M_1}_{0}|\psi^{M_1}_{p_1}}=\exp\Big(-\frac{1}{4}a^2k^2\Big)\,.
\ee
Similarly, we can compute $\alpha_3$ and find that $\alpha_3=\alpha_3^*=\alpha_1$. In real experiments, the wave length 
of the photon is much larger than the width $a$; so we have $\alpha_1=\alpha_3=\alpha_3^*\sim 1$. 
This is exactly the PI case in Fig.\ref{mw}(a).  One may want to 
use a photon with much shorter wave length so that $\alpha_1=\alpha_3=\alpha_3^*\ll 1$ and $\beta_1=\beta_3=\beta_3^*\sim 1$, i.e., 
the PS case.  However, the interaction of mirrors with shorter-wave-length 
photon is very different and the MZI can consequently cease to work.\\

There is one possible way to realize the PS case as illustrated in Fig.\ref{mw}(b). This is to add a very sensitive detector $DP$ that is
capable of measuring the tiny momentum that a mirror gains after interacting with the photon. If the momentum is zero, the detector
is described by $\ket{DP_0}$;  if the momentum is $\vec{p}_1$ or $\vec{p}_3$, the detector has the state $\ket{DP_1}$. 
These two states should be orthogonal to each other $\braket{DP_0|DP_1}=0$ to reflect the effectiveness of the detection. 
With the addition of the new detector, the interaction $U_e$ in Eq.(\ref{m1}) can be re-written as
\ba
&&U_e\ket{\phi_H}\otimes \ket{\psi^{M_1}_0}\otimes \ket{DP_0}\nonumber\\
&=&\frac{1}{\sqrt{2}}\left(\ket{\phi_H}\otimes \ket{\psi^{M_1}_0,DP_0}-\ket{\phi_V}\otimes \ket{\psi^{M_1}_{p_1},DP_1}\right)\,.\nonumber\\
\label{m1p}
\ea
Let $\ket{\widetilde{\psi}^{M_1}_0}=\ket{\psi^{M_1}_0,DP_0}$ and $\ket{\widetilde{\psi}^{M_1}_p}=\ket{\psi^{M_1}_{p_1},DP_1}$. 
We  clearly have $\braket{\widetilde{\psi}^{M_1}_0|\widetilde{\psi}^{M_1}_p}=0$. As a result, when we expand  as in Eq.(\ref{exp1})
for $\ket{\widetilde{\psi}^{M_1}_p}$, we should have $\alpha_1=0$ and $\beta_1=1$. Similarly, we should have $\alpha_3=0$ and $\beta_3=1$.
In this way, we have effectively realized the PS case in Fig.\ref{mw}(b), where the worlds have split twice with no interference. 
Note that the discussion with the detector $DP$ is a matter of principles, not for realistic realization. In real experiments, other methods 
may be used to distinguish the two states $\ket{\psi^{M_1}_0}$ and $\ket{\psi^{M_1}_{p_1}}$ or tell which direction the photon is going after 
encountering the mirror $M_1$. \\
\begin{figure}[t]
 \includegraphics[width=7.5cm]{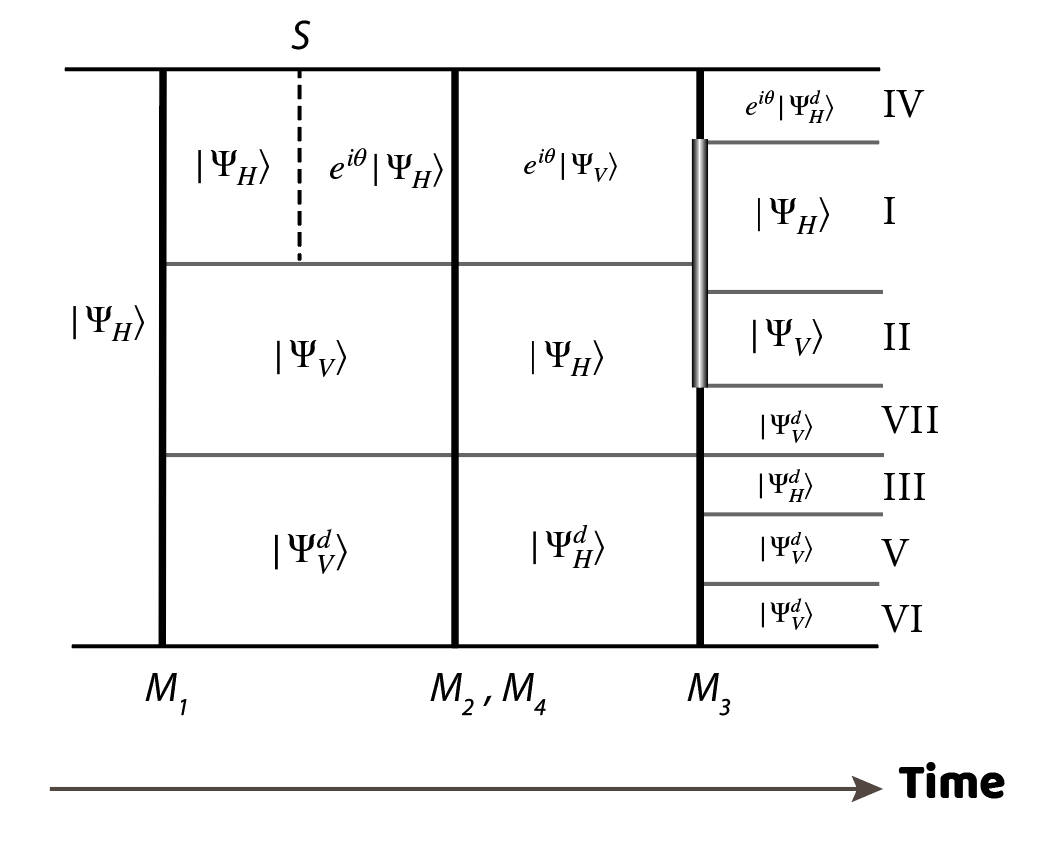}
        \caption{Worlds in the universe of Mach-Zehnder interferometer in the general case. For clarity, the weight of each world, which is 
        represented by the length of the vertical line, is not plotted to scale. The superscript $d$ in $\ket{\Psi_H^d}$ and $\ket{\Psi_V^d}$ represents the states of mirrors
         has changed relative to the initial state of the MZI universe. }
	\label{general}
\end{figure}

We now discuss the general case. As the above analysis  shows that $\alpha_1=\alpha_3=\alpha$ and $\beta_1=\beta_3=\beta^*_3=\beta$, 
we have 
\ba
\ket{\Psi_f}&=&\frac{1}{2}\Big[\alpha e^{i\frac{\theta}{2}}\cos^2\frac{\theta}{2}
\ket{\phi_H,\psi^{M_1}_0,\psi^{M_3}_0,\psi^{D_H}_1,\psi^{D_V}_0}\nonumber\\
&&+(\alpha^2-e^{i\theta})\ket{\phi_V,\psi^{M_1}_0,\psi^{M_3}_0,\psi^{D_H}_0,\psi^{D_V}_1}\nonumber\\
&&+\beta\ket{\phi_H,\psi^{M_1}_\perp,\psi^{M_3}_0,\psi^{D_H}_1,\psi^{D_V}_0}\nonumber\\
&&+\beta e^{i\theta}\ket{\phi_H,\psi^{M_1}_0,\psi^{M_3}_\perp,\psi^{D_H}_1,\psi^{D_V}_0}\nonumber\\
&&+\alpha \beta \ket{\phi_V,\psi^{M_1}_\perp,\psi^{M_3}_0,\psi^{D_H}_0,\psi^{D_V}_1}\nonumber\\
&&+\alpha \beta\ket{\phi_V,\psi^{M_1}_0,\psi^{M_3}_\perp,\psi^{D_H}_0,\psi^{D_V}_1}\nonumber\\
&&+\beta^2\ket{\phi_V,\psi^{M_1}_\perp,\psi^{M_3}_\perp,\psi^{D_H}_0,\psi^{D_V}_1}\Big]\nonumber\\
&&\otimes\ket{\psi^{M_2}_{0},\psi^{M_4}_{0}}\,.
\ea
This wave function is a superposition of seven mutually orthogonal elements, each of which describe a world. 
In the order in the above equation, we call them worlds I, II, III, IV, V, VI, and VII. 
The worlds I and II are the results of interference with 
the world I existing only in the PI case while the  world II existing in both the PI and PS cases. 
The worlds III, IV, and VII exist in the PS case. The  worlds V and VI are new and 
do not exist in either of the two special cases.  To understand these two new worlds, 
we expand the interaction in Eq.(\ref{m1}) with Eq.(\ref{exp1})
\ba
&&U_e\ket{\phi_H}\otimes \ket{\psi^{M_1}_0}=\frac{1}{\sqrt{2}}\Big[\ket{\phi_H}\otimes \ket{\psi^{M_1}_0}\nonumber\\
&&-\alpha\ket{\phi_V}\otimes \ket{\psi^{M_1}_0}-\beta\ket{\phi_V}\otimes \ket{\psi^{M_1}_\perp}\Big]\,.
\label{m1exp}
\ea
The term with $\alpha$ represents that the photon changes its direction by the mirror $M_1$ but the mirror state does not change. 
We call it reflection with no detection.  The term with $\beta$ represents that 
the photon changes its direction by the mirror $M_1$ while the mirror state becomes orthogonal to its original state.
We call it reflection with detection.  So, in the world V, the photon is reflected by the mirror $M_1$ with detection
and then reflected by the mirror $M_3$ with no detection; in the  world VI, the photon is reflected by the mirror $M_1$ with no 
detection and then reflected by the mirror $M_3$ with  detection. \\

The analysis with the general case illustrates a crucial point that the photon interferes  only when its different components, 
$\phi_H$ and  $\phi_V$, do not cause difference in the rest of the universe (e.g. the mirrors and detectors). Whenever the different
components of an object's wave function cause difference in other objects, interference disappears and decoherence occurs. \\

In his long thesis,   \E offered an insight into quantum measurement. In his view, quantum measurement is 
a generation of entanglement between two subsystems by an interacting Hamiltonian.   We now illustrate it
with the entanglement-generation interaction described in Eq.(\ref{m1exp}). The photon is the ``apparatus" 
whose reading is given by the operator $\hat{A}_p$. The eigenstates of $\hat{A}_p$ are 
\be
\ket{\phi_{\pm}}=\frac{1}{\sqrt{2}}(\ket{\phi_H}\pm\ket{\phi_V})\,,
\ee 
such that $\hat{A}_p=\pm\ket{\phi_{\pm}}$. The system is the mirror $M_1$, whose property to be measured is given by 
the operator $\hat{B}_M$.  The eigenstates of $\hat{B}_M$ are
 \be
\ket{\psi_{1,2}}=\frac{1}{\sqrt{2}}(\ket{\psi^{M_1}_0}\mp \ket{\psi^{M_1}_\perp})
\ee 
with eigenvalues being 1 and 2, respectively. Before the interaction between the photon and the mirror, we have
\be
\ket{\phi_H}\otimes \ket{\psi^{M_1}_0}=\frac{1}{2}(\ket{\phi_{+}}+\ket{\phi_{-}})\otimes (\ket{\psi_{1}}+\ket{\psi_{2}})\,,
\ee
where there is no entanglement between the photon and the mirror at all.  After the interaction $U_e$, we have Eq.(\ref{m1exp}). 
We first consider the special PS case, $\alpha=0$ and $\beta=1$. 
In this case, we can re-write the right hand side of Eq.(\ref{m1exp}) as 
\ba
&&\frac{1}{\sqrt{2}}\Big(\ket{\phi_H}\otimes \ket{\psi^{M_1}_0}
-\ket{\phi_V}\otimes \ket{\psi^{M_1}_\perp}\Big)\nonumber\\
&=&\frac{1}{\sqrt{2}}\Big(\ket{\phi_+}\otimes \ket{\psi_1}+\ket{\phi_-}\otimes \ket{\psi_2}\Big)\,.
\label{dbasis}
\ea
We have a maximum entanglement (a perfect correlation) here: when the apparatus (photon) reads `+', 
we know the mirror is in the state $\ket{\psi_1}$; when the apparatus reads `$-$', the mirror is in the state $\ket{\psi_2}$. 
In the general case, after the interaction $U_e$, we have
\ba
&&U_e\ket{\phi_H}\otimes \ket{\psi^{M_1}_0}\nonumber\\
&=&\sqrt{\frac{1-\alpha}{2}}\ket{\phi_+}\otimes\Big(\widetilde{\alpha}\ket{\psi_1}+\frac{\alpha}{\widetilde{\alpha}}\ket{\psi_2}\Big)
\nonumber\\
&+&\sqrt{\frac{1+\alpha}{2}}\ket{\phi_-}\otimes\Big(\frac{\alpha}{\widetilde{\alpha}}\ket{\psi_1}+\widetilde{\alpha}\ket{\psi_2}\Big)\,,\nonumber\\
\label{dbasis2}
\ea
where $\widetilde{\alpha}=(\sqrt{1+\alpha}+\sqrt{1-\alpha})/2$. We no longer have a perfect measurement. 
When the apparatus (photon) reads either `+' or  `$-$',  the mirror is not  in the eigenstates of  $\hat{B}_M$, the target of our measurement. 
When $\alpha$ is only slightly smaller than one, the resulted states are very close to the eigenstates of  $\hat{B}_M$ and can be regarded
as an approximation. \E call this kind of measurement approximate measurement. It is clear in the special universe of MZI 
the approximate measurement is more common than the precise measurement. It is  the same in our universe, the general universe.\\

Several caveats are warranted here. (1) The approximation is not the result of noises or other random factors in real experimental setup. 
(2) The two operators $\hat{A}_p$ and $\hat{B}_M$ are introduced for theoretical illustration; it seems unlikely that they can be 
realized in real experiments. (3) To the best of my knowledge, nobody appears to have studied approximate measurement thoroughly 
since \E, many fundamental questions need to be answered, for example, the precise definition of approximate measurement.  \\

The above discussion has led us to another intriguing issue in quantum mechanics. We use Eq. (\ref{dbasis}) as an illustration. 
On the left hand side, we have a familiar  universe that has split into two worlds: in one world, the photon moves 
horizontally and the mirror stays the same; in the other world, the photon moves vertically and the mirror changes into a state orthogonal 
to its original state. On the right hand side, the same universe is split to two very different worlds: in one world, 
the photon is in the state $\phi_+$, an  eigenstate of $\hat{A}_p$, and the mirror is  in the state $\psi_1$, an  eigenstate of $\hat{B}_p$;
in the other  world, the photon is in the state $\phi_-$ and the mirror is  in the state $\psi_2$. 
In fact, there are infinite ways to re-write this entangled state. So, which  represents the reality?  For us, 
the world where the photon moves either horizontally or vertically is the reality since we have the ability to measure the photon's position
and the ability to measure whether the mirror has momentum or not. If a different kind of creature or instrument can make
measurements according to $\hat{A}_p$ and $\hat{B}_p$, then the right hand side of Eq. (\ref{dbasis}) is the reality. \\

What kind of world that we perceive depends on our abilities of perception. These different abilities mathematically correspond to different bases. 
Suppose $\ket{\Phi_{\rm U}}$ is the wave function for the whole universe. 
For one group of observers $O_A$ with a given set of measurement abilities, it can be decomposed in a set of basis as
\be
\ket{\Phi_{\rm U}}=\ket{\Phi_1}+\ket{\Phi_2}+\cdots+\ket{\Phi_j}+\cdots
\ee
These $\ket{\Phi_j}$'s  are the worlds perceived by $O_A$. For another group of observers $O_B$ with a different set of measurement abilities, 
the universe wave function  can be decomposed in a different set of basis as
\be
\ket{\Phi_{\rm U}}=\ket{\widetilde{\Phi}_1}+\ket{\widetilde{\Phi}_2}+\cdots+\ket{\widetilde{\Phi}_j}+\cdots
\ee
The worlds $\Phi_j$'s are very different from the worlds $\ket{\widetilde{\Phi}_j}$'s. 
It is possible that even the space-time that we are experiencing may look very different for another group of observers. 

\part{Abridgment of  Everett's long thesis}
This abridgment is done by mostly paraphrasing Everett's  long thesis; Everett's words in complete sentences 
are rarely used.  Chapters in the thesis become sections in this abridgment. 
\E  used correlation or canonical correlation to mean entanglement in his thesis;  
I use entanglement  in the abridgment wherever correlation is meant entanglement in Everett's thesis. 
In some cases the mathematical notation of Everett has been updated to more modern style, such as in the use of Dirac bracket notation. 
The author's words are indicated with {\it italic font}. 
\section{Introduction}
An isolated quantum system is completely described by a wave function $\ket{\psi}$. According to standard textbooks 
on quantum mechanics
the wave function $\ket{\psi}$ can change in two fundamentally different ways~\cite{Neumann}
\begin{description}
\item[Process 1] Observation with respect to operator $\hat{O}$ that has eigenfunctions $\ket{\phi_1},\ket{\phi_2},\ket{\phi_3},\cdots$ 
 will transforms {\it discontinuously}  the wave function $\ket{\psi}$ to  one of the eigenfunctions, $\ket{\phi_j}$, with probability $|\braket{\phi_j|\psi}|^2$. 
\item[Process 2] Continuous and deterministic change of the state $\ket{\psi}$ with time according to the Schr\"odinger equation 
\be
\i\hbar\frac{\partial}{\partial t}|\psi\rangle=\hat{H}\ket{\psi}\,,
\ee
where $\hat{H}$ is the operator. 
\end{description}
Process 1 is commonly known as the collapse of wave function. \\

The above scheme can lead to a paradox when there are more than one observer. Consider a room isolated in space where one observer A is to 
perform a measurement on a system S and will record the result in a notebook. 
The observer A is aware that the system S is in a quantum state $\ket{\psi}$ that is not in an eigenstate of the measurement. 
Another observer B is outside of the room.  Beside knowing the quantum state $\ket{\psi}$ 
and A is to perform a specified measurement, B has no interaction at all with the room and everything inside the room. 
The observer A performs the measurement and records the 
result in the notebook. One week later, B enters the room and performs his measurement, that is, taking a look at the notebook. A and B 
soon find themselves disputing each other: A insists that Process 1 (the collapse of the wave function $\ket{\psi}$ occurred when 
he performed the measurement. B is confident that the whole room should evolve according to Process 2 for one week. Process 1 occurred
only when he enters the room and performs his observation by looking at the notebook. There are five different ways to resolve the paradox
or the dispute between A and B. \\
 
\begin{description}
\item[Alternative 1] To postulate that there is only one observer in the universe.
\item[Alternative 2] To limit the applicability  of quantum mechanics: quantum theory fails when it is applied to observers, measuring devices, 
or more generally any system of macroscopic size.
\item[Alternative 3] To deny  the possibility of the outside observer $B$ could ever be in possession of the state function of $A$ and $S$, where
$A$ is the observer inside the lab and $S$ is the quantum system that $A$ measures. 
\item[Alternative 4] To abandon the position that a wave function is a complete description of a system. 
\item[Alternative 5] To assume that the universal validity of the quantum description by the complete abandonment of Process 1, i.e., the collapse of wave function. 
\end{description}
 Alternatives 1 and 2 are clearly hard to defend. Alternative 4 can be viewed as hidden variable theory. {\it Local hidden variable theory 
 has been refuted by Bell's inequality~\cite{Bell}.  
Alternative 3 is a bit ambiguous, at least in my opinion.} No matter what, the first four alternatives need additional assumptions. 
In contrast, alternative 5  has many advantages:
\begin{itemize}
\item It relies on two basic ingredients of quantum mechanics: (1) The wave function $\ket{\psi}$ in a Hilbert space 
offers a complete description of a quantum system; (2)  $\ket{\psi}$ evolves unitarily according to the 
\sdg~equation.
\item The quantum theory applies to the entire universe.
\item Measurement is no longer a special process and can be described as any other physical processes.
\end{itemize}
The key for developing alternative 5 is to study composite quantum systems and exploit the entanglement ( or correlation in Everett's own words) 
between subsystems.

\section{Probability, information, and correlation}
This section (or chapter as used in Everett's thesis) offers a very general mathematical treatment of information and correlation, 
which is used in later sections to define correlation (or entanglement) between different  quantum subsystems. 
\subsection{Finite joint distribution}
For a collection of finite sets, ${\mathcal X}^1,{\mathcal X}^2,\cdots,{\mathcal X}^n$, we can define a joint probability distribution, 
$P(x^1_i,x^2_j,\cdots,x^n_k)$, where $x^1_i\in {\mathcal X}^1, x^2_j\in {\mathcal X}^2,\cdots,x^n_k\in{\mathcal X}^n$. 
This is the probability that events $x^1_i,x^2_j,\cdots,x^n_k$ occur simultaneously. We can also define the marginal distribution
\ba
&&P(x^1_i,x^2_k,\cdots,x^j_\ell)=\nonumber\\
&&\sum_{{\mathcal X}^{j+1},{\mathcal X}^{j+2},\cdots,{\mathcal X}^n}P(x^1_i,x^2_k,\cdots,x^j_\ell,x^{j+1}_p,\cdots,x^n_q)\,,\nonumber\\
\ea
where the summation is over all possible elements in ${\mathcal X}^{j+1},{\mathcal X}^{j+2},\cdots,{\mathcal X}^n$. 
This is the probability that events $x^1_i,x^2_k,\cdots,x^j_\ell$ occur with no restrictions on other sets  
${\mathcal X}^{j+1},{\mathcal X}^{j+2},\cdots,{\mathcal X}^n$. The conditional distribution is defined as
\ba
&&P_{x^{j+1}_p,\cdots,x^n_q}(x^1_i,x^2_k,\cdots,x^j_\ell)\nonumber\\
&=&\frac{P(x^1_i,x^2_k,\cdots,x^j_\ell,x^{j+1}_p,\cdots,x^n_q)}{P(x^{j+1}_p,\cdots,x^n_q)}\,,
\label{eq:cond}
\ea
which is the probability that events $x^1_i,x^2_k,\cdots,x^j_\ell$ occur while other variables are fixed at $x^{j+1}_p,\cdots,x^n_q$.\\

For any function $f(x^1_i,x^2_j,\cdots,x^n_k)$ defined on sets ${\mathcal X}^1,{\mathcal X}^2,\cdots,{\mathcal X}^n$, 
its expectation is defined as
\be
\braket{f}=\sum_{{\mathcal X}^{1},{\mathcal X}^{2},\cdots,{\mathcal X}^n}P(x^1_i,x^2_j,\cdots,x^n_k)f(x^1_i,x^2_j,\cdots,x^n_k)
\ee
where the summation is over all possible values in sets ${\mathcal X}^1,{\mathcal X}^2,\cdots,{\mathcal X}^n$. 
Two variables ${\mathcal X}^{1}$ and ${\mathcal X}^{2}$ are independent if the joint distribution $P(x^1_i,x^2_j)=P(x^1_i)P(x^2_j)$.

\subsection{Information for finite distributions}
For a single random variable ${\mathcal X}$ with distribution $P(x_i)$, its information $I_{\mathcal X}$ is defined as
\be
I_{\mathcal X}=\sum_i P(x_i)\ln P(x_i)=\braket{\ln P(x_i)}\,.
\label{eq:info}
\ee
This  is just the negative of Shannon's entropy. If ${\mathcal X}$ has $m$ different values, 
the maximum of $I_{\mathcal X}$ is zero and the minimum of $I_{\mathcal X}$ is $-\ln m$.  The former corresponds to 
the case where one value, say $x_j$, has $P(x_j)=1$ and the other values have $P(x\neq x_j)=0$. The latter is the case
where every value has the same probability $P(x_j)=1/m$. This definition can be easily generalized for many variables
${\mathcal X}^1,{\mathcal X}^2,\cdots,{\mathcal X}^n$
\ba
&&I_{{\mathcal X}^1,{\mathcal X}^2,\cdots,{\mathcal X}^n}\nonumber\\
&=&\sum_{{\mathcal X}^1,{\mathcal X}^2,\cdots,{\mathcal X}^n}
P(x^1_i,x^2_j,\cdots,x^n_k)\ln P(x^1_i,x^2_j,\cdots,x^n_k)\nonumber\\
&=&\sum_{{\mathcal X}^1,{\mathcal X}^2,\cdots,{\mathcal X}^n}
\braket{\ln P(x^1_i,x^2_j,\cdots,x^n_k)}\,.
\ea
Similarly, one can also define information for the conditional distribution.
It is clear that if all the random variables ${\mathcal X}^1,{\mathcal X}^2,\cdots,{\mathcal X}^n$ are independent from each other, 
we have 
\be
I_{{\mathcal X}^1,{\mathcal X}^2,\cdots,{\mathcal X}^n}=I_{{\mathcal X}^1}+I_{{\mathcal X}^2}+\cdots+I_{{\mathcal X}^n}\,.
\ee
\subsection{Correlation for finite distributions}
For two random variables ${\mathcal X}$ and ${\mathcal Y}$, the correlation between them is defined as 
\be
C({\mathcal X}, {\mathcal Y})=I_{{\mathcal X},{\mathcal Y}}-I_{{\mathcal X}}-I_{{\mathcal Y}}\,.
\label{eq:corr}
\ee
It is clear that we have $C({\mathcal X}, {\mathcal Y})=0$ if two random variables ${\mathcal X}$ and ${\mathcal Y}$ are independent. 
This definition can be generalized to group correlations. Suppose we have groups of random variables, 
${\mathcal X}^1,{\mathcal X}^2,\cdots,{\mathcal X}^n$; ${\mathcal Y}^1,{\mathcal Y}^2,\cdots,{\mathcal Y}^m$; $\cdots$;
${\mathcal Z}^1,{\mathcal Z}^2,\cdots,{\mathcal Z}^\ell$, the correlation between these groups is
\ba
&&C({\mathcal X}^1,\cdots,{\mathcal X}^n; {\mathcal Y}^1,\cdots,{\mathcal Y}^m; \cdots;
{\mathcal Z}^1,\cdots,{\mathcal Z}^\ell)\nonumber\\
&=&I_{{\mathcal X}^1,\cdots,{\mathcal X}^n; {\mathcal Y}^1,\cdots,{\mathcal Y}^m; \cdots;
{\mathcal Z}^1,\cdots,{\mathcal Z}^\ell}\nonumber\\
&&-I_{{\mathcal X}^1,\cdots,{\mathcal X}^n}-
I_{{\mathcal Y}^1,\cdots,{\mathcal Y}^m}-\cdots-I_{{\mathcal Z}^1,\cdots,{\mathcal Z}^\ell}\,.
\ea
A special case of this group correlation is 
\ba
&&C({\mathcal X}^1,{\mathcal X}^2,\cdots,{\mathcal X}^n)
=I_{{\mathcal X}^1,{\mathcal X}^2,\cdots,{\mathcal X}^n}\nonumber\\
&&-I_{{\mathcal X}^1}-I_{{\mathcal X}^2}-\cdots-I_{{\mathcal X}^n}\,.
\ea
\subsection{Generalization and further properties of correlation}
We shall now generalize the definition of correlation to joint probability distributions over arbitrary sets of any cardinality.
To do this,  we consider the refinement of a finite distribution. Consider a random variable ${\mathcal X}$ consisting 
of finite number of events $\{x_1,x_2,\cdots,x_n\}$. It is possible that the event $x_i$ is actually the disjunction 
of several exclusive events $\{\tilde{x}_{i,1},\tilde{x}_{i,2},\cdots,\tilde{x}_{i,\mu}\}$.
The distribution $P^\prime (\tilde{x}_{i,\nu})$ is called a refinement of the distribution $P(x_i)$
\be
P(x_i)=\sum_{\nu}P^\prime (\tilde{x}_{i,\nu})\,,
\ee
where the summation is over all possible values of $\nu$ for a given $x_i$.   This can easily be generalized to multiple variables. 
For a distribution $P(x_i,y_j)$ of two random variables  ${\mathcal X}$ and ${\mathcal Y}$ 
and its refinement $P(\tilde{x}_{i,\mu},\tilde{y}_{j,\nu})$, there exist two correlations 
$C({\mathcal X},{\mathcal Y})$ and $C^\prime({\mathcal X},{\mathcal Y})$, respectively. 
There is an interesting and important relation between these two corrections 
\be
C^\prime({\mathcal X},{\mathcal Y})\ge C({\mathcal X},{\mathcal Y})\,.
\ee
With this relation, we can generalize the correlation to any probability measure over continuous variables. \\

For simplicity, we consider two 
continuous random variables ${\mathcal X}$ and ${\mathcal Y}$ and a probability 
measure $M({\mathcal X},{\mathcal Y})$ over their cartesian product. 
We can divide ${\mathcal X}$ into finite subsets ${\mathcal X}_i$ and ${\mathcal Y}$ into finite subsets ${\mathcal Y}_j$.
This naturally leads to a probability distribution $P_1({\mathcal X}_i,{\mathcal Y}_j)$, which can be obtained  by integration
of $M({\mathcal X},{\mathcal Y})$ over these subsets. With $P_1({\mathcal X}_i,{\mathcal Y}_j)$, we can compute the 
correlation $C_1({\mathcal X},{\mathcal Y})$ between ${\mathcal X}$ and ${\mathcal Y}$.  By further dividing 
the subsets ${\mathcal X}_i$ and ${\mathcal Y}_j$, we can have another correlation $C_2({\mathcal X},{\mathcal Y})$.
By repeating the process, we have a sequence of correlations
\be
C_1({\mathcal X},{\mathcal Y})\le C_2({\mathcal X},{\mathcal Y})\le \cdots \le C_n({\mathcal X},{\mathcal Y})\,.
\ee
As a result, the correlation between two continuous random variables ${\mathcal X}$ and ${\mathcal Y}$ is defined as
\be
C({\mathcal X},{\mathcal Y})=\lim_{n\rightarrow\infty}C_n({\mathcal X},{\mathcal Y})\,,
\ee
where $n\rightarrow\infty$ means that the division becomes finer and finer, approaching the continuous limit. \\

Suppose that $f$ is a one-one map, ${\mathcal U}=f({\mathcal X})$, and 
$g$ is a one-one map, ${\mathcal V}=g({\mathcal Y})$. We have
\be
C({\mathcal X},{\mathcal Y})=C({\mathcal U},{\mathcal V})\,.
\ee
This shows that the correlation is invariant under one-to-one transformation. 

\subsection{Information for general distribution}
For a random variable ${\mathcal X}$ with a finite set of values $\{x_i\}$,  we assign a positive number $a_i$ to each value $x_i$. 
These $a_i$ are called information measure. If the probability distribution is $P(x_i)$, its information relative to this information measure
is defined as
\be
I_X=\sum_i P(x_i)\ln \frac{P(x_i)}{a_i}=\braket{\ln \frac{P(x_i)}{a_i}}\,.
\ee
For multiple variables, say, ${\mathcal X},{\mathcal Y},{\mathcal Z}$, with information measures $\{a_i\}$, $\{b_j\}$, $\{c_k\}$, 
respectively, and a joint probability distribution $P(x_i,y_j,z_k)$, their information relative to these measures are 
\be
I_{XYZ}=\sum_{ijk}P(x_i,y_j,z_k)\ln\frac{P(x_i,y_j,z_k)}{a_ib_jc_k}\,.
\ee
The previous definition of information is a special case where all values of $\{a_i\}$, $\{b_j\}$, $\{c_k\}$ in the information 
measure are unity.  Interestingly, the correlation $C({\mathcal X},{\mathcal Y},{\mathcal Z})$ is independent of information measure. \\

The advantage of introducing information measure is that we can now generalize information for continuous variables. For example, 
for a continuous variable, ${\mathcal X}$, with a probability distribution $P(x)$, we can divide it into finite sets ${\mathcal X}_i$ and use $\mu_i$ for the Lebesgue measure of the set ${\mathcal X}_i$. We then have 
\be
I^\mu_X=\sum_i P(x_i)\ln \frac{P(x_i)}{\mu_i}\,.
\ee
where $P(x_i)$ is the probability over the set ${\mathcal X}_i$. We can further divide and refine the sets ${\mathcal X}_i$ and 
define informations correspondingly. These informations form a series which has an upper limit. We define this upper limit as
the information for ${\mathcal X}$ with probability distribution $P(x)$
\be
I_X=\lim_\mu I^\mu_X\,.
\ee

\section{Quantum mechanics}
Quantum mechanics has two basic ingredients: (1) the states of a quantum system are vectors in a Hilbert space;
(2) the time evolution of an isolated quantum system is given by a linear wave equation. 
{\it One crucial question is whether we need more to relate quantum mechanics to our experimental and daily experience.
Many physicists represented by von Neumann think that we need at least one more ingredient, Process 1, which 
was mentioned at the beginning. \E thinks that no more ingredient (or assumption) is needed.}

\subsection{Composite quantum systems}
\def\mn{\mathcal N}
Consider a pair of quantum systems $S_1$ and $S_2$.  If their Hilbert spaces are $\mh_1$ and $\mh_2$, respectively, 
the Hilbert space of the composite system $S=S_1+S_2$ is $\mh_1\otimes \mh_2$. 
If $\ket{\xi_j}$ is a complete orthonormal set for  $\mh_1$ and $\ket{\eta_j}$ for $\mh_2$, a general state of
$S=S_1+S_2$  can be expressed as
\be
\ket{\Psi}=\sum_{ij}c_{ij}\ket{\xi_i,\eta_j}\,,
\ee
where $\ket{\xi_i,\eta_j}$ is a shorthand for $\ket{\xi_i}\otimes\ket{\eta_j}$. The concepts introduced 
in the last section can be applied here. 
Let $\hat{A}$ be a Hermitian operator on $S_1$ with eigenfunctions $\ket{\phi_i}$ and  eigenvalues $\mu_i$
and $\hat{B}$ be a Hermitian operator on $S_2$ with eigenfunctions $\ket{\varphi_i}$ and  eigenvalues $\nu_i$. Then
\be
P_{ij}=|\braket{\phi_i,\varphi_j|\Psi}|^2
\ee
is a joint {\bf square-amplitude} distribution of the quantum state $\ket{\Psi}$ over $\hat{A}$ and $\hat{B}$. 
{\it Note that \E did not use probability distribution here. The physical meaning of  {\bf square-amplitude} is discussed 
later.} It has two marginal distributions
\be
P(\phi_i)=\sum_j|\braket{\phi_i,\varphi_j|\Psi}|^2\,,
\ee
and 
\be
P(\varphi_j)=\sum_i|\braket{\phi_i,\varphi_j|\Psi}|^2\,.
\ee
Correspondingly, there are two conditional distributions 
\be
P_j(\phi_i)=P_{ij}/P(\varphi_j)\,,
\ee
\be
P_i(\varphi_j)=P_{ij}/P(\phi_i)\,.
\ee
These distributions can be used to compute the marginal and conditional expectations of $\hat{A}$ or  $\hat{B}$. \\

A key concept introduced by \E is relative state. 
For a given state $\ket{\eta}$ in $S_2$, 
there is a corresponding relative state in $S_1$, 
\be
\ket{\phi_{\eta}}=\mn_\eta\braket{\eta|\Psi}=\mn_\eta\sum_{ij}c_{ij}\ket{\xi_i}\braket{\eta|\eta_j}\,,
\ee
where $\mn_\eta$ is a normalization constant. For a given state $\ket{\eta}$, its relative state $\ket{\phi_{\eta}}$ is clearly unique and independent of $\ket{\xi_i}$ and $\ket{\eta_i}$. 
The relative state $\ket{\phi_{\eta}}$ can be used to compute expectation of any operator $\hat{A}$ on $S_1$ 
conditioned by the state $\ket{\eta}$ in $S_2$. \\

If $\ket{\eta}$ is one of the basis states $\ket{\eta_j}$, we have 
\be
\ket{\phi_{\eta_j}}=\mn_{\eta_j}\sum_{i}c_{ij}\ket{\xi_i}\,.
\ee
It is clear that 
\be
\ket{\Psi}=\sum_j \frac{1}{\mn_{\eta_j}}\ket{\phi_{\eta_j}}\otimes\ket{\eta_j}\,.
\ee
Two different relative states $\ket{\phi_{\eta_j}}$ and $\ket{\phi_{\eta_k}}$ are not necessarily orthogonal 
\be
\label{nonrel}
\braket{\phi_{\eta_j}|\phi_{\eta_k}}=\mn_{\eta_j}\mn_{\eta_k}\sum_{i}c^*_{ij}c_{ik}\neq 0\,.
\ee

In a general state $\ket{\Psi}$ of $S$, the subsystem $S_1$ can not be described by a single state but by a mixture of states. 
We usually use the density matrix to describe this kind of mixture. For the whole system, it is always in a pure state $\ket{\Psi}$
and its density matrix is
\be
\rho=\ket{\Psi}\bra{\Psi}\,.
\ee
By tracing out the subsystem $S_2$, we have the density matrix for the subsystem $S_1$

\be
\rho_1=\sum_{ik}\sum_jc_{ij}c_{kj}^*\ket{\xi_i}\bra{\xi_k}=\sum_j  \frac{1}{\mn_{\eta_j}^2}\ket{\phi_{\eta_j}}\bra{\phi_{\eta_j}}\,.
\ee 
Similarly,  we can define $\rho_2$. The most important conclusion of this section  
is that it is meaningless to ask the absolute state of a subsystem -- 
one can only ask the state relative to a given state of the remainder of the system. 
{\it What \E is discussing here is of course entanglement: in a composite system where the subsystems 
are entangled, the subsystems are described by density matrices not pure quantum states}.  

 \subsection{Information and correlation in quantum mechanics}
 Consider an operator $\hat{A}$, which has eigenstates $\ket{\xi_i}$ with eigenvalues $\mu_i$.  
 The information of this operator 
 in a given state $\ket{\psi}$ is defined as
 \be
 I_{A}(\psi)=\sum_i |\braket{\xi_i|\psi}|^2\ln |\braket{\xi_i|\psi}|^2\,.
 \ee
The operator $\hat{A}$ has been assumed to be non-degenerate. If $\hat{A}$  is degenerate, 
that is, for eigenvalue $\mu_i$, there are multiple eigenstates $\ket{\xi_{i,\lambda}}~(\lambda=1,2,\cdots,m_i)$, 
its information is defined as
\be
 I_{A}(\psi)=\sum_i \left(\sum_\lambda |\braket{\xi_{i,\lambda}|\psi}|^2\right)\ln 
 \frac{\sum_\lambda|\braket{\xi_{i,\lambda}|\psi}|^2}{m_i}\,.
 \ee

For convenience, we introduce projection operator
\be
{\rm P}_i=\sum_\lambda\ket{\xi_{i,\lambda}}\bra{\xi_{i,\lambda}}\,,
\ee
with which we have
\be
\sum_\lambda|\braket{\xi_{i,\lambda}|\psi}|^2={\rm Tr}(\rho {\rm P}_i)\,,
\ee
where $\rho=\ket{\psi}\bra{\psi}$.
The information of operator $\hat{A}$ now has a concise form
\be
 I_{A}(\psi)=\sum_i {\rm Tr}(\rho {\rm P}_i)\ln \frac{{\rm Tr}(\rho {\rm P}_i)}{m_i}\,.
 \ee

We consider again the composite system $S=S_1+S_2$. For the operator $\hat{A}$ that acts only on $S_1$ 
and $\hat{B}$ only on $S_2$,  we assume for simplicity that both of them have  no degeneracy. 
Their joint information is 
\be
I_{A,B}(\Psi)=\sum_{ij} {\rm Tr}(\rho {\rm P}_i^{A}{\rm P}_j^{B})
\ln {\rm Tr}(\rho {\rm P}_i^{A}{\rm P}_j^{B})\,,
\ee
where ${\rm P}_i^{A}=\ket{\xi_i}\bra{\xi_i}$, ${\rm P}_j^{B}=\ket{\eta_j}\bra{\eta_j}$, and 
$\rho=\ket{\Psi}\bra{\Psi}$. The marginal informations for the operator $\hat{A}$ 
and the operator $\hat{B}$ are
\be
I_{A}(\Psi)=\sum_i {\rm Tr}(\rho_1 {\rm P}_i^{A})\ln {\rm Tr}(\rho_1 {\rm P}_i^{A})
\ee
and 
\be
I_{B}(\Psi)=\sum_j {\rm Tr}(\rho_2 {\rm P}_j^{B})\ln {\rm Tr}(\rho_2 {\rm P}_j^{B})\,,
\ee
where $\rho_1={\rm Tr}_2(\ket{\Psi}\bra{\Psi})$ and $\rho_2={\rm Tr}_1(\ket{\Psi}\bra{\Psi})$. 
We define the correlation between ${\hat A}$ and ${\hat B}$ as
\ba
&&C_{A,B}(\Psi)=I_{A,B}(\Psi)-I_{A}(\Psi)-I_{B}(\Psi)\nonumber\\
&=&\sum_{ij}{\rm Tr}(\rho {\rm P}_i^{A}{\rm P}_j^{B})
\ln \frac{{\rm Tr}(\rho {\rm P}_i^{A}{\rm P}_j^{B})}{{\rm Tr}(\rho_1 {\rm P}_i^{A}){\rm Tr}(\rho_2 {\rm P}_j^{B})}\,.~~~~
\ea
~\\

Without loss of generality, we assume that the dimension of the Hilbert space $\mh_1$ is equal 
or bigger than $\mh_2$.  
The reduced density matrix $\rho_2$ for the subsystem $S_2$ 
 is a Hermitian matrix and it can be diagonalized with non-negative eigenvalues. Suppose that 
its eigenvectors are $\ket{\varphi_j}$ with eigenvalues $v_j$. If   $\ket{\xi_i}$ is the 
complete basis of the subsystem $S_1$,  the relative state of $\ket{\varphi_j}$ for a 
general state $\ket{\Psi}$ is 
 \be
 \ket{\phi_{j}}=\mn_{\varphi_j}\sum_{i}\ket{\xi_i}\braket{\xi_i,\varphi_{j}|\Psi}\,.
 \ee
 One can show that $\ket{\phi_{j}}$'s are orthonormal to each other,
 \ba
 \braket{\phi_{j}|\phi_{k}}&=&\mn_{\varphi_j}\mn_{\varphi_k}\sum_{i,l}
 \braket{\xi_i|\xi_l}\braket{\xi_i,\varphi_{j}|\Psi}\braket{\Psi|\xi_l,\varphi_{k}}\nonumber\\
 &=&\mn_{\varphi_j}\mn_{\varphi_k}\sum_{i}
 \braket{\xi_i,\varphi_{j}|\Psi}\braket{\Psi|\xi_i,\varphi_{k}}\nonumber\\
 &=&\mn_{\varphi_j}\mn_{\varphi_k}\sum_{i}\braket{\varphi_{j}|\rho_2|\varphi_{k}}=\delta_{j,k}\,.
 \ea
 Note the difference between here and Eq.(\ref{nonrel}), where $\ket{\eta_j}$'s are not the eigenstates of $\rho_2$. 
  As a result, we have
 \be
 \ket{\Psi}=\sum_j v_j \ket{\phi_j,\varphi_j}\,.
 \ee
 {\it This  is called canonical representation of $\ket{\Psi}$ by \E and it is of course just the Schmidt decomposition.} 
 Now let us choose $\hat{A}=\rho_1$ and $\hat{B}=\rho_2$, and assume that  
 there is no degeneracy in eigenvalues $v_j$. We have
 \be
  I_{\rho_1,\rho_2}(\Psi)=I_{\rho_1}(\Psi)=I_{\rho_2}(\Psi)=\sum_j v_j\ln v_j
 \ee
 Consequently, we have
 \ba
 &&C_{\rho_1,\rho_2}(\Psi)=-\sum_j v_j\ln v_j\nonumber\\
 &=&-{\rm Tr}(\rho_1\ln\rho_1)=-{\rm Tr}(\rho_2\ln\rho_2)\,.
 \ea
 This special correlation is called canonical correlation by \E and it is, of course, exactly entanglement. 
 For convenience, we let ${\mathcal C}(\Psi)=C_{\rho_1,\rho_2}(\Psi)$.  
One may conjecture that,  for any pair of operator ${\hat A}$ on $S_1$ and operator ${\hat B}$ on $S_2$, 
 the following inequality holds
 \be
 C_{{\rm A},{\rm B}}(\Psi)\le {\mathcal C}(\Psi)\,.
 \ee
 {\it This conjecture has now been proved rigorously by Donald}~\cite{Donald}.\\
 
 For operators $\hat{x}$ and $\hat{k}=\hat{p}/\hbar$, there is the Heisenberg uncertainty relation 
 \be
 \Delta x\Delta k\ge \frac{1}{4}\,.
 \ee
\E conjectured that in terms of information this relation can be written as
 \be
 I_x+I_k\le \ln\frac{1}{\pi e}\,.
 \ee
 {\it This conjecture has been proved in Ref.~\cite{BM}, where \E was  not acknowledged. }
 
\subsection{Measurement}
{\it \E regarded measurement as a natural process in quantum mechanics and there is no fundamental 
distinction between ``measuring apparatus" and other physical 
systems.  For \E, a measurement is simply a special interacting process between two quantum subsystems, 
which results in the end  that the property of the measured subsystem is correlated to a quantity 
in the measuring subsystem. The measuring process has two characteristics that 
distinguish it from other interacting processes.} \\

Suppose that we have two subsystems $S_1$ and $S_2$, initially in a product state $\ket{\Psi_0}=\ket{\psi_0,\phi_0}=\ket{\psi}\otimes\ket{\phi}$. 
The system will evolve dynamically under a Hamiltonian $\hat{H}$ of the whole system. According to the analysis in the above subsections, 
at any moment, the overall state $\ket{\Psi(t)}$ can be decomposed canonically as
\be
\ket{\Psi(t)}=\sum_j p_j\ket{\psi_j(t),\phi_j(t)}
\ee
where $\ket{\psi_j(t)}$'s and $\ket{\phi_j(t)}$'s are eigenfunctions of two operators $\hat{A}(t)$ and $\hat{B}(t)$, respectively. 
The Hamiltonian $\hat{H}$ is said to generate a measurement if the following limits exist
\be
\hat{A}_\infty=\lim_{t\rightarrow \infty}\hat{A}(t)~~,~~~~~~\hat{B}_\infty=\lim_{t\rightarrow \infty}\hat{B}(t)
\ee 
and they do not depend on initial conditions. \\

There is one requirement for a Hamiltonian $\hat{H}$  to generate a measurement: $\hat{H}$ does not decrease the information in the marginal
distribution of $\hat{A}$. This means that if initially $\ket{\Psi_0}=\ket{\zeta,\phi_0}$ where $\ket{\zeta}$ is an eigenfunction of $\hat{A}$, 
we should have at any time that $\ket{\Psi(t)}=\ket{\zeta,\phi(t)}$. The requirement is necessary for
the repeatability of measurements: if a spin is measured to be up along the $z$ direction, it should be 
still up when we measure it again along the $z$ direction.\\

In sum, a Hamiltonian $\hat{H}$ is said to generate a measurement of $\hat{A}$ 
in $S_1$ by $\hat{B}$ in $S_2$ if the following two conditions are satisfied: (1) the correlation $C_{A,B}$ increases to its maximum with time;
(2) $\hat{H}$ does not decrease  the marginal information of $\hat{A}$.\\

We now turn to a model proposed by von Neumann~\cite{Neumann} to illustrate the 
above definition of quantum measurement.  This model consists of a particle of  one 
coordinate $\hat{q}$ and an apparatus of one coordinate $\hat{r}$ (which may represents the position of a meter needle). 
The interaction between them is very strong so that we neglect all the kinetic energies. 
This means that the whole Hamiltonian is given by 
\be
\hat{H}_I=-i\hbar q\frac{\partial }{\partial r}\,.
\ee
If the initial condition is a product state 
\ba
\ket{\Psi_0}&=&\int dq\phi(q)\ket{q}\int dr\eta(r)\ket{r}\nonumber\\
&=&\int dqdr\phi(q)\eta(r)\ket{q,r}\,,
\ea
it is straightforward to find the evolution of this state
\be
\ket{\Psi(t)}=\int dqdr\phi(q)\eta(r-qt)\ket{q,r}\,.
\ee
Let us consider a special case $\eta(r)=\delta(r-r_0)$,  that is, 
the apparatus needle initially points to a definite position $r_0$.  In this case, we have
\be
\ket{\Psi(t)}=\int dq\phi(q)\ket{q,r_0+qt}\,.
\label{superp}
\ee
It is clear that $\hat{H}_I$ has kept the marginal information of $\hat{q}$. Let us consider the correlation $C_{q,r}(t)$. 
Initially, we have $C_{q,r}(t)=0$. At time $t$, we have
\ba
&&C_{q,r}(t)=I_{q,r}(t)-I_q(t)-I_r(t)\nonumber\\
&=&-I_q(0)=-\int dq |\phi(q)|^2\ln  |\phi(q)|^2\,.
\ea
The correlation $C_{q,r}(t)$ has increased to its maximum as soon as $t$ is not zero. The above analysis 
clearly shows that the Hamiltonian $\hat{H}_I$ generates a measurement of $\hat{q}$ 
for the system by $\hat{r}$ of the apparatus. {\it The general case that the apparatus needle has 
no definite position initially is more complicated and 
was discussed in the long thesis by \E.}\\


In the above discussion,  the apparatus initially has a definite position $r_0$. After measurement, 
 the apparatus no longer has a definite position. In fact, according to Eq.(\ref{superp}), the apparatus 
 is in a superposition of states of different positions and  the probability of
 its position at $r_0+qt$ is $|\phi(q)|^2$.  If this apparatus is  of macroscopic size, 
 this means that its meter needle does not point to a definite position. 
 We of course have never seen this kind of measurement  in any laboratory or similar phenomena 
in our daily life. {\it To resolve this dilemma, one possible way to assume that
the mysterious collapse of wave function (Process 1) during the measurement. \E found that one can 
resolve  this dilemma  within the framework of quantum mechanics without 
additional assumption.}  \\
 
 
\section{Observation}
Observers are introduced as purely physical systems and are treated completely 
within the framework of quantum mechanics. In other words, observers are simply usual 
quantum systems. If this treatment is successful, 
it should  build a consistent picture between the appearance of phenomena, i.e., 
the subjective experience of observers, and the usual probabilistic interpretation of quantum mechanics.  
\subsection{Formulation of the problem}
One can regard an observer as an automatically functioning machine that has sensors and
the capacity to record or register past sensory data and machine configurations.  
When an observer $O$ has observed the event $\alpha$, it means that $O$ has changed to a new state that depends on $\alpha$. 
Observers are assumed to have memories; the subjective experience of an observer is 
related to the contents of its memory. As a result,  the quantum state of an observer $O$ should be written as
\be
\ket{\psi^O_{[A,B,\cdots,C]}}
\ee
where $A, B,\cdots,C$ represent memories in the order of time. Sometimes $[\cdots,A, B,\cdots,C]$ is 
used to indicate  the possible previous memories 
that are not relevant for the current observations. \\

Consider  an observer $O$ who wants to measure (or observe) the property $\hat{A}$  of a system $S$. 
The eigenfunctions of $\hat{A}$ are $\ket{\phi_j}$'s.  
Initially, the system $S$ is in one of the eigenfunctions $\ket{\phi_j}$'s of $\hat{A}$ and the observer is in state $\ket{\psi^O_{[\cdots]}}$.  
A good observation is defined as the one that results in transforming 
\be
\ket{\psi^{S+O}}=\ket{\phi_j}\otimes \ket{\psi^O_{[\cdots]}}=\ket{\phi_j;\psi^O_{[\cdots]}}
\ee
to 
\be
\ket{\tilde{\psi}^{S+O}}=\ket{\phi_j}\otimes \ket{\psi^O_{[\cdots,\alpha_j]}}=\ket{\phi_j;\psi^O_{[\cdots,\alpha_j]}}
\ee
The semicolon is used here and will be used to delimit the system state and the observer state. 
The requirement that the system state $\ket{\phi_j}$ is unchanged is necessary if you want the observation is repeatable. 
It is clear that observation is just  quantum measurement (introduced in the last section) with memories. 

\subsection{Deductions}
If the system initially is in a general quantum state described by $\sum_j a_j\ket{\phi_j}$, 
the final total state after a good observation is
\be
\ket{\tilde{\psi}^{S+O}}=\sum_j a_j \ket{\phi_j;\psi^O_{[\cdots,\alpha_j]}}\,.
\label{eq:supero}
\ee
This follows directly from the superposition principle and is consistent 
with the general framework of quantum mechanics. Two features stand out in the above equations.
(1) There is entanglement between the system and the observer and, as a result, neither 
of them has its independent state. (2) 
The result seems to contradict our daily experience.  On the one hand, 
the final states are superposition of many different states, 
each of which corresponds to a definite observation outcome; on the other hand, 
there is only one outcome in our daily experience. \\


{\it Here comes Everett's genius. \E thinks that each superposition element in Eq.(\ref{eq:supero}) 
represents a ``world" and the observer observes different outcomes in different  ``worlds". 
Since the quantum dynamical evolution is linear, which respects the  superposition, 
each world evolves on its own and in each world the observer experiences only 
one definite outcome.  This is in accordance with our daily experience; 
at the same time, no additional assumption, such as the collapse of wave function, is needed.
This is the so-called many-worlds interpretation. However, \E himself never
called each superposition element  ``world"; ``many-worlds" was coined 
by de Witt in 1970s~\cite{DeWitt}. \E call it branch.} \\

The above observation should be the same even in the presence of other systems 
which do not interact with the observer $O$. We thus have the general rules of observation.\\

{\bf Rule 1} - The observation of a quantity $\hat{A}$, with eigenfunctions $\ket{\phi}^{S_1}_j$, 
in a system $S_1$ by the observer $O$, transforms the total state according to 
\ba
&&\ket{\psi^{S_1},\psi^{S_2},\cdots,\psi^{S_n};\psi^O_{[\cdots]}}\nonumber\\
&\rightarrow& \sum_j a_j \ket{\phi^{S_1}_j,\psi^{S_2},\cdots,\psi^{S_n};\psi^O_{[\cdots,\alpha_j]}}\,,
\ea
where $\ket{\psi^{S_1}},\ket{\psi^{S_2}},\cdots,\ket{\psi^{S_n}}$ are the initial quantum states for systems $S_1,S_2,\cdots,S_n$, respectively, and 
$a_j=\braket{\phi^{S_1}_j|\psi^{S_1}}$.\\

{\bf Rule 2} - {\bf Rule 1} may be applied separately to each element of a superposition of total system states, 
the results are superposed to obtain the final total state. 
Thus, a determination of $\hat{B}$, with eigenfunctions $\ket{\phi_k}^{S_2}$, on $S_2$ by the observer $O$ transforms the total state
\be
 \sum_j a_j \ket{\phi^{S_1}_j,\psi^{S_2},\cdots,\psi^{S_n};\psi^O_{[\cdots,\alpha_j]}}
\ee
to 
\be
 \sum_{jk} a_j b_k\ket{\phi^{S_1}_j,\phi^{S_2}_k,\psi^{S_3},\cdots,\psi^{S_n};\psi^O_{[\cdots,\alpha_j,\beta_k]}}\,,
\ee
where $b_k=\braket{\phi^{S_2}_k|\psi^{S_2}}$. These two rules follow directly 
from the superposition principle and are consistent 
with the general framework of quantum mechanics. \\

Consider again the simple case  where there is one system 
and one observer. The observation results in Eq.(\ref{eq:supero}). 
If one repeats this observation, according to {\bf Rule 2}, the total state becomes
\be
\sum_j a_j \ket{\phi_j;\psi^O_{[\cdots,\alpha_j,\alpha_j]}}\,.
\ee
Each superposition element in the above now describes that the observer has obtained the 
same result for both observations. That is, in each world, the observation is repeatable. 
This is consistent with our experience. \\

Let us go one step further by considering many different systems which are initially in the same state
\be
\ket{\psi^{S_1}}=\ket{\psi^{S_2}}=\cdots=\ket{\psi^{S_n}}=\sum_j a_j \ket{\phi_j}\,.
\ee
Therefore, the initial state of the total system is 
\be
\ket{\psi^{S_1+S_2+\cdots+S_n+O}_0}=\ket{\psi^{S_1},\psi^{S_2},\cdots,\psi^{S_n};\psi^O_{[\cdots]}}\,.
\ee
The measurement is performed on the systems in the order $S_1,S_2,\cdots,S_n$. 
After the first measurement on $S_1$, we have 
\be
\ket{\psi^{S_1+S_2+\cdots+S_n+O}_1}=
\sum_j a_j\ket{\phi_j^1,\psi^{S_2},\cdots,\psi^{S_n};\psi^O_{[\cdots,\alpha_j^1]}}\,.
\ee
The total state after the second measurement is
\ba
&&\ket{\psi^{S_1+S_2+\cdots+S_n+O}_2}=\nonumber\\
&&\sum_j a_ja_k\ket{\phi_j^1,\phi_k^{2},\psi^{S_3},\cdots,\psi^{S_n};\psi^O_{[\cdots,\alpha_j^1,\alpha_k^2]}}\,.
\ea
After $r\le n$ measurements have taken place, we have
\ba
&&\ket{\psi^{S_1+S_2+\cdots+S_n+O}_r}=\sum_j a_ja_k\cdots a_\ell\nonumber\\
&&\ket{\phi_j^1,\phi_k^{2},\cdots,\phi_\ell^{r}; \psi^{S_{r+1}},\cdots,\psi^{S_n};
\psi^O_{[\cdots,\alpha_j^1,\alpha_k^2,\cdots,\alpha_\ell^r]}}\,.\nonumber\\
\ea
Each of the superposition elements, which is one of the many worlds,  describes an observer which has observed an apparently random sequence of definite results represented  by $[\cdots,\alpha_j^1,\alpha_k^2,\cdots,\alpha_\ell^r]$. If one repeats the measurement on the system $S^m$ $(m<r)$, 
the observer would get a memory sequence of $[\cdots,\alpha_j^1,\cdots,\alpha_k^m,\cdots,\alpha_\ell^r,\alpha_k^m]$. In each world, the observer
feels the ``collapse" of wave function. \\

To make sense of the coefficients $a_ja_k\cdots a_\ell$ before each superposition element, we need to assign a measure to them. We first consider a
superposition state
\be
\ket{\psi}=\sum_j a_j \ket{\phi_j}\,.
\ee 
To assign a measure, we first need that each element is normalized $\braket{\phi_j|\phi_j}=1$. 
For each element $\ket{\phi_j}$, the assigned measure is $\mathcal M(a_j)$, which is a non-negative function.  We could change $\ket{\phi_j}$ 
to $e^{i\theta}\ket{\phi_j}$ and $a_j$ to $e^{-i\theta}a_j$, then the measure $\mathcal M(a_j)$ assigned to this element becomes 
$\mathcal M(e^{-i\theta}a_j)$. However, physically nothing has been changed. Therefore, we need 
$\mathcal M(a_j)=\mathcal M(e^{-i\theta}a_j)$. For this to be true, it is clear that $\mathcal M(a_j)$
should depend only on the amplitude of $a_j$, that is, $\mathcal M(|a_j|)$. \\

In reality, due to the accuracy of the measurement or other reasons, we often regard a group of states 
$\ket{\phi_p},\ket{\phi_{p+1}},\cdots, \ket{\phi_q}$ as the same, i.e., 
\be
\tilde{a}\ket{\tilde{\phi}}=\sum_{j=p}^{q}a_j\ket{\phi_j}\,,
\ee 
where $\braket{\tilde{\phi}|\tilde{\phi}}=1$ is normalized. We require the additivity for the measure, that is, 
\be
{\mathcal M}(|\tilde{a}|)=\sum_{j=p}^{q} \mathcal M(|a_j|)\,.
\ee
$\braket{\tilde{\phi}|\tilde{\phi}}=1$ implies that 
\be
|\tilde{a}|^2=\sum_{j=p}^{q} |a_j|^2
\ee
and 
\be
{\mathcal M}(\sqrt{\sum_{j=p}^{q} |a_j|^2})=\sum_{j=p}^{q}{\mathcal M}(|a_j|)\,.
\ee
The only choice is the {\bf square amplitude} measure, ${\mathcal M}(|a_j|)=c|a_j|^2$, where $c$ can 
be fixed by requiring $\sum_j{\mathcal M}(|a_j|)=1$. \\

This square amplitude measure ${\mathcal M}$ is of probability nature. {\it \E discussed this for general cases. 
I'll use a simple case to illustrate. Let us consider a simple  system of spin-1/2. Suppose that there are $n$ copies of them  
and their states are the same
\be
\ket{\psi^{S_1}}=\ket{\psi^{S_2}}=\cdots=\ket{\psi^{S_n}}=\frac{1}{\sqrt{2}}( \ket{u}+\ket{d})\,,
\ee 
where $\ket{u}$ is for spin up and  $\ket{d}$ is for spin down. We make observations of the spins of $\hat{\sigma}^z$.  
If the spin up $\ket{u}$ is registered as 0 and 
the spin down $\ket{d}$ is registered as 1, we have after measuring all the spins
\ba
&&\ket{\psi^{\rm final}_n}=\frac{1}{\sqrt{2^n}}\sum \nonumber\\
&&\ket{u^1,d^{2},\cdots,d^{r}u^{r+1},\cdots,u^{n};
\psi^O_{[\cdots,0^1,1^2,\cdots,1^{r}0^{r+1},\cdots,0^{n}]}}\,,\nonumber\\
\ea
where the summation is over all possible sequences of 0's and 1's of length $n$. 
Most of times, we only care about how many of these spins are up and how many of them  down. So, we group these 
states  according to how many spins are up
\be
\sqrt{\frac{n!}{m!(n-m)!2^n}}\ket{\tilde{\phi}^m}=\frac{1}{\sqrt{2^n}}\sum_{m~{\rm ups}}\ket{u\cdots d}\,.
\ee
where $\ket{\tilde{\phi}^m}$ represents a state where $m$ of the $n$ spins are up. The measure ${\mathcal M}$ 
for the group of $m$ up spins is
\be
{\mathcal M}_m=\frac{n!}{m!(n-m)!2^n}\,,
\ee
which is exactly the probability that one observes the state of spin up $m$ times when making $n$ repeated same measurements. 
What happens here is that, after $n$ measurements, it splits into $2^n$ branches of worlds, each of which is equally probable and 
has a different sequence of 0's and 1's registered. The chance being in a world where there are $m$ up spins (or $m$ 0s) is ${\mathcal M}_m$. 
In general, if the spin is in a state $\ket{\phi}=a\ket{u}+b\ket{d}$, we still have $2^n$ branches of worlds after  $n$ measurements but 
the chance being in a world where there are $m$ up spins (or $m$ 0's) is 
\be
{\mathcal M}_m^\prime=\frac{n!}{m!(n-m)!}|a|^{2m}|b|^{2(n-m)}\,.
\ee
The number of branches has nothing to do with the probability measure $|a|^2$ or $|b|^2$; it depends on the observation outcomes.} \\


The above results can be straightforwardly generalized to the cases where different measurements are performed on different systems 
and different measurements are performed on the same system. 

\subsection{Several observers}
{\it It was pointed out at the beginning that the assumption of Process 1 (or the collapse of wave function) would lead to 
self-contradiction when there are more than one observers. There is no such contradiction in the many-worlds theory. }
Let us consider the situation where there are multiple observers. Three different cases are to be considered.\\

 {\bf Case 1}: Two observers observe the same quantity in the same system.\\

Observers $O_1$ and $O_2$  are to observe the quantity $\hat{A}$ for the system $S$ that is in the following state
\be
\ket{\psi^S}=\sum_j a_j\ket{\phi_j}\,,
\ee
where $\ket{\phi_j}$ is an eigenstate of $\hat{A}$. The observer $O_1$ makes the first observation; 
we apply {\bf Rule 1} to the initial state 
$\ket{\Psi_0}=\ket{\psi^S; \psi^{O_1}_{[\cdots]},\psi^{O_2}_{[\cdots]}}$ and obtain
\be
\ket{\Psi_1}=\sum_j a_j\ket{\phi_j; \psi^{O_1}_{[\cdots,\alpha_j]},\psi^{O_2}_{[\cdots]}}\,.
\ee
The observer $O_2$ makes the second observation; we apply {\bf Rule 2} and obtain
\ba
\ket{\Psi_2}&=&\sum_{j,k} a_j\braket{\phi_k|\phi_j}\ket{\phi_j; \psi^{O_1}_{[\cdots,\alpha_j]},\psi^{O_2}_{[\cdots,\alpha_k]}}\nonumber\\
&=&\sum_{j} a_j\ket{\phi_j; \psi^{O_1}_{[\cdots,\alpha_j]},\psi^{O_2}_{[\cdots,\alpha_j]}}\,.
\ea
This shows that the first observation by $O_1$ leads to splitting of different branches of worlds, and  the second observation by $O_2$ 
of the same quantity causes no splitting and furthermore $O_2$ observes the same result as $O_1$.  It is in accordance 
with our daily experience: two observers measuring the same quantity on a given system always obtains the same result. 
{\it This result can clearly be generalized to any number of observers.} \\

{\it \E even considered the situation where the two observers are allowed to communicate their observation results. 
It does not lead to any self-contradiction and contradiction to our daily experience. }\\

{\it Let us now return to the paradox in Section I. The observer A made the measurement; the observer B did not make the measurement 
directly and he obtained his result by reading A's notebook. In this case, we have 
$\sum_{j} a_j\ket{\phi_j; \psi^{A}_{[\cdots,\alpha_j]},\psi^{B}_{[\cdots,\alpha_{j,A}]}}$, where the subscript 
indicates that the result comes from $A$. In each world, the two observers A and B always agree with each other and 
have nothing to argue about. The paradox is resolved.}\\

{\bf Case 2}: Two observers measure separately two different quantities, which are non-commuting, in the same system.\\

The same initial state $\ket{\Psi_0}$ and the same observation by $O_1$. Then the observer $O_2$ measures the quantity $\hat{B}$, which 
does not commute with $\hat{A}$. We apply  {\bf Rule 2} to $\ket{\Psi_1}$ and obtain
\be
\ket{\widetilde{\Psi}_2}=\sum_{j,k} a_j\braket{\varphi_k|\phi_j}\ket{\varphi_j; \psi^{O_1}_{[\cdots,\alpha_j]},\psi^{O_2}_{[\cdots,\beta_k]}}\,,
\ee
where $\ket{\varphi_k}$ is the eigenstate of $\hat{B}$. In this case, the second observation leads to further splitting. If $\hat{A}$ has $N_A$ 
eigenstates and $\hat{B}$ has $N_B$ eigenstates, then there are $N_AN_B$ different worlds in total, which are represented 
by the terms on the left hand side of the above equation. The measure ${\mathcal M}$ of 
the coefficients $a_j\braket{\varphi_k|\phi_j}$ gives the probability of getting into one of the worlds 
if the observations are repeated on the same system in the same state.\\

{\bf Case 3}: Two observers $O_1$ and $O_2$ measure two  entangled systems $S_1$ and $S_2$: $O_1$ measures
$\hat{A}$ in $S_1$  and $O_2$ measures $\hat{B}$ in $S_2$. \\

For simplicity, we assume that the initial state of the composite system of $S_1$ and $S_2$ is entangled 
\be
\ket{\psi^{S_1+S_2}}=\sum_{j=1}^{N} a_j \ket{\phi_j,\varphi_j}\,,
\label{entangled}
\ee
where $N\le N_A, N_B$.  There is no interaction between $S_1$ and $S_2$ during the following observations. 
The total initial state is 
\be
\ket{\Psi_0}^\prime=\ket{\psi^{S_1+S_2}; \psi^{O_1}_{[\cdots]},\psi^{O_2}_{[\cdots]}}\,.
\ee 
After $O_1$ observes $\hat{A}$ in $S_1$, the total state becomes
\be
\ket{\Psi_1}^\prime=\sum_{j=1}^{N} a_j\ket{\phi_j,\varphi_j; \psi^{O_1}_{[\cdots,\alpha_j]},\psi^{O_2}_{[\cdots]}}\,.
\ee
There are now $N$ different branches of worlds. The observer $O_2$ then observes $\hat{B}$ in $S_2$ and transforms the 
total state to
\be
\ket{\Psi_2}^\prime=\sum_{j=1}^{N} a_j\ket{\phi_j,\varphi_j; \psi^{O_1}_{[\cdots,\alpha_j]},\psi^{O_2}_{[\cdots,\beta_j]}}\,.
\ee
No more splitting and there are still $N$ different branches of worlds. In each branch, when $O_1$ observes the result 
represented by $\ket{\phi_j}$,  $O_2$ observes the result represented by $\ket{\varphi_j}$.  The observation results of 
$O_1$ and $O_2$ are correlated.  It is easy to check that if $O_2$ observes first and $O_1$ second, the end state is still
$\ket{\Psi_2}^\prime$.  It is clear that the observations of $O_1$ and $O_2$ do not influence each other. Furthermore, 
if $O_1$ repeats its measurement of $\hat{A}$ in $S_1$,  then the total state $\ket{\Psi_2}^\prime$ is turned into
\be
\ket{\Psi_2}^\prime=\sum_{j=1}^{N} a_j\ket{\phi_j,\varphi_j; \psi^{O_1}_{[\cdots,\alpha_j,\alpha_j]},\psi^{O_2}_{[\cdots,\beta_j]}}\,.
\ee
We would have the same total state if $O_1$ had observed $\hat{A}$ in $S_1$ twice in a row before $O_2$ observed $\hat{B}$ in $S_2$. 
This shows that in every world $O_1$ would not know whether $O_2$ has made an observation on $S_2$ or not if there is no direct
communication between them. In other words, the entanglement in the state (\ref{entangled}) can not be used for communication. 

\section{Supplementary topics}
We have presented an abstract treatment of measurement and observation completely 
within the framework of quantum mechanics, which are in correspondence with our experience. 
{\it Upon observation and measurement, there is splitting into different worlds: in each world to an observer 
there appears  a collapse of wave function (or Process 1); however, with all the worlds combined, the evolution is always unitary. 
This approach has at least three advantages: (1) it is logically self-consistent; (2) it does not involve any additional assumption, for example,
the collapse of wave function; (3) it is completely quantum mechanical with no use of classical concepts. }

\subsection{Macroscopic objects and classical mechanics}
In the many-worlds theory, there is no more divide between quantum world and classical world. Macroscopic objects are also described 
by wave functions. However, we do experience in our daily life a classical world where macroscopic objects have definite positions and momenta, 
moving around according to classical mechanics. Below is a rough explanation how the classical world emerges from quantum mechanics with no
detailed proof. \\ 

Let us first consider a simple case, the hydrogen atom. Its wave function is essentially a product of a centroid wave function and a wave function
for the relative coordinate between the proton and the electron. The former describes the motion of the hydrogen atom as a whole in space and time;
the latter is usually a bound state that gives us the size and shape of the hydrogen atom. The situation is similar for macroscopic object that we see 
daily. For example, the wave function of a cannonball can be written roughly as
\be
\ket{\psi_c}=\ket{g(X)}\otimes\ket{\psi_X^b}=\ket{g(X),\psi_X^b}\,,
\ee
where $\ket{g(X)}$ is a  Gaussian wave function well localized at the position $X$ and $\ket{\psi_X^b}$  is the bound state
giving us the size and shape of a cannonball located at $X$. When the centroid wave function $\ket{g(X)}$ evolves over a long period of time, 
it can spread to occupy a large region of space. Therefore, in the general case, the cannonball is not necessarily well localized 
and its state is given by
\be
\ket{\psi_c^\prime}=\int \ket{a(X)}\otimes\ket{\psi_X^b}dX\,,
\ee
where $\ket{a(X)}$ is any smooth and normalizable function.  {\it In this general state, there is an entanglement between the centroid position 
 $X$ and the rest of the coordinates of a cannonball. \\
 
 However, in most of systems, we can separate the centroid coordinates
from the coordinates for the relative motion in the Schr\"odinger equation; as a result, $\ket{\psi_X^b}$ does not depend on $X$ explicitly.
In this case, we can take $\ket{\psi_X^b}$ outside of the above integral and have $\ket{\psi_c^\prime}=\ket{a}\otimes\ket{\psi^b}$.
 When $\ket{a}$ is not well localized, the state $\ket{\Psi_c^\prime}$ 
represents a kind of  ``smeared out" cannonball.  In contrast, we only see cannonballs with definite positions and momenta in our daily life.  
This dilemma can be resolved by noticing that cannonballs are never truly isolated and they are constantly be observed or measured by photons
and other objects. We assume that someone magically set up a cannonball in a superposition state of two well separated and localized Gaussians, that is, 
\be
\ket{\widetilde{\psi}_c}=\frac{1}{\sqrt{2}}\big(\ket{g(X_1)}+\ket{g(X_2)}\big)\otimes\ket{\psi^b}\,.
\ee
The initial state of this cannonball and its observer is $\ket{\Psi^{c+O}_0}=\ket{\widetilde{\psi}_c;\psi^O_{[\cdots]}}$, where  
$\ket{\psi^O_{[\cdots]}}$ is the state of an observer, which can be photons or other physical objects that can distinguish the difference 
between the positions $X_1$ and $X_2$.  After the interaction between the cannonball and the observer (which happens in a very short time), 
a splitting of worlds happens and the total state becomes
\ba
\ket{\Psi^{c+O}_1}&=&\frac{1}{\sqrt{2}}\big(\ket{g(X_1);\psi^O_{[\cdots,X_1]}}\nonumber\\
&&+\ket{g(X_2);\psi^O_{[\cdots,X_2]}}\big)\otimes\ket{\psi^b}\,.
\label{cball}
\ea
In one world, the observer finds the cannonball well localized at $X_1$; in the other world, the observer finds the cannonball well localized at $X_2$.
This is why we do not observe  ``smeared out" cannonballs.  The cannonball and other similar macroscopic objects with well localized wave functions
will move approximately according to the classical mechanics. After a certain period of time, the wave packet will spread out so much to cause another 
splitting of worlds. The detailed account of how long a well-localized wave packet will follow the classical trajectory is given by Ehrenfest time~\cite{Ehrenfest}.
\E did not use the concept of Ehrenfest time in his thesis. }\\

\subsection{Ampilication processes}
In our abstract discussion of measuring process in the previous sections, we have simplified the coupling between the system and the observer (or the apparatus). 
In reality, there is a chain of intervening systems linking a microscopic system to a macroscopic apparatus. Each system in the chain of intervening systems is
correlated to its predecessor, resulting an amplification of effects from the microscopic system to a macroscopic apparatus. \\

We use Geiger counter as an example to illustrate this amplification process. A Geiger counter contains a large number of gas atoms that are placed in a strong
electric field. The atoms are metastable against ionization. More importantly, the product of ionizing one gas atom can cause ionization of more atoms in a cascading 
process. This chain reaction correlates large number of gas atoms: either very few or very many of the gas atoms are ionized at a given time. \\

To put the above discussion in a mathematical form, we write the state of a Geiger counter in terms of its individual gas atoms
\be
\ket{\psi^G}=\sum_{ij\cdots k}a_{ij\cdots k}\ket{\phi_i,\phi_j,\cdots,\phi_k}\,,
\ee
where $\ket{\phi_i,\phi_j,\cdots,\phi_k}$ represents a state where the first atom is in the $i$th state, the second atom is in the $j$th state, ..., the $n$th atom 
is in the $k$th state. The superposition terms on the right hand side of the above equations describe either large number of 
 ionized atoms or few  ionized atoms.   Due to the chain ionization, there are almost no terms for medium-sized number of ionized atoms. 
 By choosing a medium-sized number, we can place these superposition terms in two groups
 \be
 a_1\ket{\psi_{[U]}}=\sum_{ij\cdots k}\!{}^\prime a_{ij\cdots k}\ket{\phi_i,\phi_j,\cdots,\phi_k}
 \ee
and 
 \be
 a_2\ket{\psi_{[D]}}=\sum_{ij\cdots k}\!{}^{\prime\prime} a_{ij\cdots k}\ket{\phi_i,\phi_j,\cdots,\phi_k}\,.
 \ee
The primed summation is over all terms with few number of ionized atoms and the double primed summation is over all terms with very large number 
of ionized atoms. $\ket{\psi_{[D]}}$ and $\ket{\psi_{[U]}}$ represent, respectively,  two macroscopic distinguishable states of 
a Geiger counter, discharged or undischarged.  As  a result,  the state of a Geiger counter can be simply  written as
\be
\ket{\psi^G}=a_1\ket{\psi_{[U]}}+a_2\ket{\psi_{[D]}}\,.
\ee
Consider a particle which is detectable by a Geiger counter. The total initial state is 
\be
\ket{\Psi^{p+G}_0}=\ket{\psi^p;\psi_{[U]}}\,,
\ee
where $\ket{\psi^p}$ is the state of the particle. If the wave function $\ket{\psi^p}$ is not well localized so that it has a part $\ket{\psi^p_o}$ 
outside of the Geiger counter and the other part $\ket{\psi^p_i}$ inside the Geiger counter, i.e., 
$\ket{\psi^p}=a\ket{\psi^p_o}+b\ket{\psi^p_i}$. After the particle encounters the Geiger counter, the total state is transformed to
\be
\ket{\Psi^{p+G}_1}=a\ket{\tilde\psi^p_o;\psi_{[U]}}+b\ket{\tilde\psi^p_i;\psi_{[D]}}\,.
\ee
{\it We have a splitting into two worlds: in one world the Geiger counter is discharged and in the other one the counter is undischarged. 
This is similar to the splitting in Eq.(\ref{cball}).}

\subsection{Reversibility and irreversibility}
In the usual treatment of quantum mechanics, there are both Process 1 (the collapse of wave function) and Process 2 (unitary evolution). 
It is obvious that Process 1 is irreversible and  Process 2 is reversible. This difference can be quantified by introducing another information
\be
I_{\rho}={\rm Tr}(\hat{\rho}\ln\hat{\rho})\,,
\ee
where $\hat{\rho}$ is a density matrix of a quantum system.  If the system changes according to Process 2, we have 
$\hat{\rho}^\prime=U\hat{\rho} U^\dagger$, which does not change $I_{\rho}$ since
\ba
I_{\rho^\prime}&=&{\rm Tr}\big(\hat{\rho}^\prime\ln\hat{\rho}^\prime\big)={\rm Tr}\big[U\hat{\rho}U^\dagger\ln(U\hat{\rho}U^\dagger)\big]\nonumber\\
&=&{\rm Tr}\big[U\hat{\rho}\ln \hat{\rho}U^\dagger\big]={\rm Tr}\big(\hat{\rho}\ln\hat{\rho}\big)=I_\rho\,.
\ea
For Process 1, we consider a simple case where the system is in a pure quantum state, that is, 
$\hat{\rho}=\ket{\psi}\bra{\psi}$. The measurement is for the quantity $\hat{A}$, whose eigenfunctions are $\ket{\phi_j}$. 
After the measurement (Process 1), there is a probability of $|\braket{\phi_j|\psi}|^2$ of the measured result is $\ket{\phi_j}$. This means that
the density matrix becomes
\be
\hat{\rho}^\prime=\sum_j |\braket{\phi_j|\psi}|^2\ket{\phi_j}\bra{\phi_j}\,,
\ee
and 
\be
I_{\rho^\prime}=\sum_j \Big(|\braket{\phi_j|\psi}|^2\ln\Big(|\braket{\phi_j|\psi}|^2\Big)\le I_{\rho}=0\,.
\ee
So, Process 1 decreases the information $I_{\rho}$ but never increase it. One can prove rigorously that this is true for any $\hat{\rho}$ not just for pure states. \\

In the many-worlds theory, even though only Process 2 is recognized, an observer can still feel similar irreversibility on the subjective level. 
When an observation is performed, it leads to a superposition of many different worlds. From this time forward, since the unitary evolution is linear, 
these worlds are parallel, evolve independently, and no longer influence each other. The observer in each world has only information in his world, 
knowing nothing about other parallel worlds. As a result, for an observer in a given world, this process is also irreversible  since he can not in principle 
get to know the state before the measurement based on the information available in his world. 
This irreversibility implies that there is a fundamental limit on the knowledge of the entire universe. \\

However, the irreversibility discussed here appears not related to the second law of thermodynamics, which reflects a different kind of irreversibility. 
{\it There are two ways to see the difference.  First,  the former is of quantum nature while the latter  is also valid in classical systems. 
When one mixes two piles of sand of different colors, there is clearly no quantum process involved but the mixing is irreversible as dictated 
by the  second law. Second, when the universe keeps splitting into more and more worlds, more and more systems get entangled together. 
In this sense, the irreversibility associated with the world splitting is for an open system whereas the second law 
of thermodynamics is for a closed system. These are definitely not the final words on the relation between the second law of thermodynamics
and the splitting of worlds, which warrants further study. In fact, Tegmark discussed the second law of thermodynamics
within the framework of many-worlds theory using a tripartite partition of the universe~\cite{Tegmark}. 
My personal view is that  to discuss the second law within quantum mechanics one has to follow von Neumann, 
who defined quantum entropy for pure states (different from the well-known von Neumann entropy) and proved quantum H-theorm 
~\cite{Neumann1929,von2010proof,Han}. }

\subsection{Approximate measurement}
In many situations, we have only approximate measurements, where the apparatus or observer interacts weakly with the system and for a finite time. 
It is hard  to understand these cases with Process 1, which requires that all measurements result in a precise projection to an eigenstate of 
a measured quantity. The position measurement appears to be the best example to illustrate this difficulty. \\

In any situation, we do not know the precise position of any particle. One possible way to understand this with Process 1 is that 
the measurement indeed results in a precise position but the observer has only imprecise information. This view is clearly wrong. In practice, for example, 
when tracking high-energy particles with cloud chambers, we can measure  the approximate positions of a particle successively.  
This means that we can predict approximately the position of a particle with its current approximate position. 
If Process 1 were true,  after the measurement, the particle would be in an eigenstate of position and its momentum would be too uncertain to 
make any meaningful prediction for its future position. This contradicts well-established experimental facts. {\it \E has offered more detailed analysis
along this line and pointed out the inadequacy of Process 1 in approximate measurement.}


\subsection{Discussion of a spin measurement example}
Consider the $z$ component of a spin-1/2 with the Stern-Gerlach setup.  In this measurement, a particle of spin-1/2 passes through 
a magnetic field that is inhomogeneous along the $z$ direction.  The measurement is essentially to couple the spin and the orbital of the same particle. 
For simplicity, we keep only the coupling part of the Hamiltonian and approximate only the constant and linear part of the inhomogeneous field
\be
\hat{H}_I\approx \mu \hat{\sigma}_z(B_0+zB_1)\,,
\ee
where $\mu$ is the magnetic moment of the particle. The initial state of the particle is assumed to be 
\be
\ket{\Psi_0}=\phi_0(z)\Big(c_1\ket{u}+c_2\ket{d}\Big)\,,
\label{sg0}
\ee
where $\phi_0(z)$ describes a wave packet along the $z$ direction and $\ket{u}$ ($\ket{d}$) is the eigenfunction of $\hat{\sigma}_z$ with eigenvalue 1 (-1). 
One can solve the Schr\"odinger equation. If  $\Delta t$ is the time  that the particle takes to traverse the field, we have 
\ba
\ket{\Psi(\Delta t)}&=&\phi_0(z)\Big[c_1e^{-i\mu (B_0+zB_1)\Delta t/\hbar)}\ket{u}\nonumber\\
&&+c_2e^{i\mu (B_0+zB_1)\Delta t/\hbar)}\ket{d}\Big]\,.
\label{sga}
\ea
This is an entangled state between the spin and the orbital. The wave function has split into two: one with momentum $\mu H_1 \Delta t$ and the other 
with momentum $-\mu H_1 \Delta t$. With long enough flying time, these two parts will become well separated in space: the upper wave packet 
for the spin up state $\ket{u}$ and  the lower wave packet for the spin down state $\ket{d}$.  The measuring ``apparatus" here is the orbital degree
of freedom of the particle, {\it which by all means is microscopic}. 

In many situations, one can regard states for composite systems such as Eq.(\ref{sga}) 
as a non-interfering mixture of states by ignoring phases in superposition elements.  For example,  it is correct when calculating 
marginal expectations for subsystems. For the state (\ref{sga}), it is alright to regard it as a mixture if one cares either the spin or the orbital but 
not both. The phase relations between different superposition elements are important. For this Stern-Gerlach system, it is possible to 
re-combine the two terms in Eq.(\ref{sga})  in another imhomogeneous magnetic field and restore the original state in Eq.(\ref{sg0})\cite{Bohm}.  
For this to happen, one can not disregard the phases. 

\section{Discussion}
We have shown that our theory ({\it the many-worlds theory}) can be put in a satisfactory correspondence with experience, and gives us 
a complete conceptual model of the universe with more than one observers. In this theory,  the wave function 
is a basic description of physical systems, including observers, and the probabilistic assertion of quantum mechanics
can be deduced from this theory as subjective appearances to the observers. This theory constitutes an objective framework 
in which puzzling subjects, such as classical phenomena, the measuring process, the inter-relationship of several 
observers, reversibility and irreversibility,  can be investigated in a logically consistent manner.  \\
 
{\it  In light of his new theory, \E discussed in length other interpretations of quantum mechanics existing  at his time.} They are 
 \begin{enumerate}[label=\alph*.]
\item {\it The ``popular" interpretation.} The wave function $\ket{\psi}$ changes continuously and deterministically with a wave equation
when the system is isolated but changes probabilistically and abruptly upon observation. 
\item {\it The Copenhagen interpretation.}  The wave function $\ket{\psi}$ is regarded as just a mathematical artifice which one uses 
to make statistical predictions. All statements about microscopic phenomena are  meaningful only within a classical experiment setup. 
\item  {\it The ``hidden variable" interpretation.} The wave function $\ket{\psi}$ is not a complete description of a system. There are
additional hidden parameters in the correct and complete theory that is to be developed in the future.
The probability in quantum mechanics is the result of our ignorance of these hidden variables. 
\item {\it The stochastic process interpretation.}  In this theory, physical systems are undergoing probabilistic changes at  all times. 
The discontinuous and probabilistic ``quantum jump" are not the result of observation and measurement but are fundamental
to the systems themselves. 
\end{enumerate}

\section{Appendices}

{\it In Everett's long thesis, there are two appendices. In the first one, \E offered detailed proofs for many mathematical relations in the main text. 
In the second one, he offered his view on theoretical physics in general. Here is the summary of the second appendix. }\\

There are a number of interpretations of quantum mechanics, most of which are equivalent in the sense that they agree with all the 
physical experiments. To decide among them, we must go beyond experiments and discuss the fundamental nature and purpose 
of physical theories. \\

Every theory has two separate parts, the formal part and the interpretive part. The formal part consists of a purely logico-mathematical structure
that consists of a collection of symbols and rules for their manipulation. The interpretive part is a set of association rules that relate
the formal symbols with the experienced world.  There can be many different theories which are logical consistent and 
correct in explaining the perceived world. In this case, further criteria such as usefulness, simplicity, comprehensiveness, pictorability, etc., must
be used to select the theory or the theories. In particular, simplicity refers to conceptual simplicity not ease in use. 
It is harmful to the progress of physics that a physical theory should contain no elements which do not correspond directly to what we observe. 

\part{The legacy of Everett's theory}
Although Everett published his short thesis in Review of Modern Physics, a well known and respected journal, along with Wheeler's supporting article~\cite{Everett,Wheeler}, his theory of the universal wave function had received little attention for many years~\cite{Byrne,Freire}. 
In 1962, \E was invited by Podolsky to a small workshop at Xavier University, where he lectured on his theory for the first time in public. 
However, this did not mean that his theory was getting wide recognition; it only showed that his work was not completely forgotten~\cite{Byrne}. \\

Everett's theory began to be noticed widely in the physics community 
only after DeWitt started to promote it as  the many-worlds theory around 1970s with Graham's help~\cite{Byrne,Freire,DeWitt}.  
As a result, Everett's theory is now widely  known as the many-worlds theory.  Only a limited number of specialists know it
as the theory of the universal wave function or the ``relative state" formulation of quantum mechanics. Now Everett's theory 
has to be mentioned in all serious discussion of  the subject known as the interpretation of quantum mechanics~\cite{Jammer}.\\

To celebrate the 50th anniversary  of its publication in 2007,  the influential magazine, Nature,  put the many-worlds theory on its July cover; 
and BBC produced a special program called ``Parallel worlds, Parallel lives".  This theory is now extremely popular among non-specialists. \\

Among physicists, the many-worlds theory is still a minority view but it is gaining momentum. 
Hawking~\cite{Hawking} and  Gell-Mann~\cite{Gellman} were
famous names among its early supporters. Current influential advocates include David Deutsch~\cite{Deutsch}, Max Tegmark~\cite{Tegmark}, and Sean Carroll~\cite{Carroll}. 
 
\acknowledgements
This work is supported by the The National Key R\&D Program of China (Grants No.~2017YFA0303302, No.~2018YFA0305602), 
National Natural Science Foundation of China (Grant No. 11921005), and 
Shanghai Municipal Science and Technology Major Project (Grant No.2019SHZDZX01).


\end{document}